\documentclass[letterpaper,twocolumn,10pt]{article}
\usepackage{zhanggroup}
\usepackage[absolute]{textpos}
\titleformat{\section}{\Large\bfseries}{\thesection}{0.8em}{}

\usepackage{epsfig,xspace}
\usepackage{amsmath,amsfonts}
\usepackage{multirow}
\usepackage{enumitem}

\newcommand{\tuple}[1]{\ensuremath{\langle #1 \rangle}}

\newcommand{\mypara}[1]{\noindent\textbf{#1.}\xspace}
\newcommand{\etal}{\textit{et al.}\xspace}
\newcommand{\ie}{\textit{i.e.}\xspace}
\newcommand{\eg}{\textit{e.g.}\xspace}

\renewcommand{\Pr}[1]{\ensuremath{\mathsf{Pr}\left[#1\right]}\xspace}

\newcommand{\gin}{\ensuremath{\mathsf{GIN}}\xspace}
\newcommand{\sage}{\ensuremath{\mathsf{SAGE}}\xspace}
\newcommand{\gat}{\ensuremath{\mathsf{GAT}}\xspace}
\newcommand{\gcn}{\ensuremath{\mathsf{GCN}}\xspace}

\newcommand{\diffpool}{\ensuremath{\mathsf{DiffPool}}\xspace}
\newcommand{\meanpool}{\ensuremath{\mathsf{MeanPool}}\xspace}
\newcommand{\mincutpool}{\ensuremath{\mathsf{MinCutPool}}\xspace}

\newcommand{\rws}{\ensuremath{\mathsf{Random Walk}}\xspace}
\newcommand{\sbs}{\ensuremath{\mathsf{Snowball}}\xspace}
\newcommand{\ffs}{\ensuremath{\mathsf{Fire Forest}}\xspace}

\newcommand{\aux}{\ensuremath{aux}\xspace}
\newcommand{\dset}{\ensuremath{\mathcal{D}}\xspace}
\newcommand{\daux}{\ensuremath{\dset_{\aux}}\xspace}

\newcommand{\nodeset}{\ensuremath{\mathcal{V}}\xspace}

\newcommand{\adj}{\ensuremath{A}\xspace}
\newcommand{\feat}{\ensuremath{X}\xspace}
\newcommand{\neigh}[1]{\ensuremath{\mathcal{N}_{#1}}\xspace}
\newcommand{\layer}{\ensuremath{\ell}\xspace}

\newcommand{\graph}{\ensuremath{\mathcal{G}}\xspace}
\newcommand{\targetgraph}{\ensuremath{\graph_T}\xspace}
\newcommand{\subgraph}{\ensuremath{\graph_S}\xspace}
\newcommand{\recongraph}{\ensuremath{\graph_R}\xspace}
\newcommand{\auxgraph}{\ensuremath{\graph_{\aux}}\xspace}

\newcommand{\embed}{\ensuremath{H}\xspace}
\newcommand{\targetembed}{\ensuremath{\embed_{\graph_T}}\xspace}
\newcommand{\subembed}{\ensuremath{\embed_{\graph_S}}\xspace}
\newcommand{\auxembed}{\ensuremath{\embed_{\graph_{\aux}}}\xspace}

\newcommand{\model}{\ensuremath{\mathcal{F}}\xspace}
\newcommand{\targetmodel}{\ensuremath{\model_T}\xspace}
\newcommand{\attackmodel}{\ensuremath{\model_A}\xspace}
\newcommand{\propinfermodel}{\ensuremath{\model_{AP}}\xspace}
\newcommand{\subinfermodel}{\ensuremath{\model_{AS}}\xspace}
\newcommand{\graphreconmodel}{\ensuremath{\model_{AR}}\xspace}
\newcommand{\targetmodelname}{target embedding model\xspace}

\newcommand{\property}{\ensuremath{\mathbb{P}}\xspace}
\newcommand{\aggr}{\ensuremath{\Phi}\xspace}
\newcommand{\upd}{\ensuremath{\Psi}\xspace}
\newcommand{\pool}{\ensuremath{\Sigma}\xspace}
\newcommand{\aggregate}[2]{\ensuremath{\aggr^{#1}\left( #2 \right)}\xspace}
\newcommand{\update}[2]{\ensuremath{\upd^{#1}\left( #2 \right)}\xspace}
\newcommand{\pooling}[1]{\ensuremath{\pool\left( #1 \right)}\xspace}
\newcommand{\messagee}{\ensuremath{\mathbf{m}}\xspace}
\newcommand{\assign}{\ensuremath{S}\xspace}
\newcommand{\loss}{\ensuremath{\mathcal{L}}\xspace}
\newcommand{\subfeat}{\ensuremath{\chi}\xspace}

\newcommand{\Lapp}[1]{\ensuremath{\mathsf{Lap}\left(#1\right)}\xspace}

\begin{document}

\begin{textblock}{12}(2,1)
\centering
To Appear in the 31st USENIX Security Symposium, August 10–12, 2022.
\end{textblock}

\title{\Large \bf Inference Attacks Against Graph Neural Networks}

\date{}

\author{
{\rm Zhikun Zhang\textsuperscript{1}\textsuperscript{\textcolor{blue!60!green}{$\ast$}}}
\ \ \
{\rm Min Chen\textsuperscript{1}\thanks{Zhikun and Min contributed equally to the paper.}}
\ \ \ \ \
{\rm Michael Backes\textsuperscript{1}}
\ \ \ 
{\rm Yun Shen\textsuperscript{2}}
\ \ \
{\rm Yang Zhang\textsuperscript{1}}
\\ 
\\
{\rm \textsuperscript{1}\textit{CISPA Helmholtz Center for Information Security}} \ \ \ 
\\
{\rm \textsuperscript{2}\textit{Norton Research Group}} \ \ \
}

\maketitle

\begin{abstract}
Graph is an important data representation ubiquitously existing in the real world. 
However, analyzing the graph data is computationally difficult due to its non-Euclidean nature.
Graph embedding is a powerful tool to solve the graph analytics problem by transforming the graph data into low-dimensional vectors. 
These vectors could also be shared with third parties to gain additional insights of what is behind the data. 
While sharing graph embedding is intriguing, the associated privacy risks are unexplored.
In this paper, we systematically investigate the information leakage of the graph embedding by mounting three inference attacks.
First, we can successfully infer basic graph properties, such as the number of nodes, the number of edges, and graph density, of the target graph with up to 0.89 accuracy.
Second, given a subgraph of interest and the graph embedding, we can determine with high confidence that whether the subgraph is contained in the target graph.
For instance, we achieve 0.98 attack AUC on the DD dataset.
Third, we propose a novel graph reconstruction attack that can reconstruct a graph that has similar graph structural statistics to the target graph.
We further propose an effective defense mechanism based on graph embedding perturbation to mitigate the inference attacks without noticeable performance degradation for graph classification tasks.\footnote{Our code is available at \url{https://github.com/Zhangzhk0819/GNN-Embedding-Leaks}.}
\end{abstract}

\section{Introduction}
\label{sec:introduction}
Many real-world systems can be represented as graphs, such as social networks~\cite{QTMDWT18}, financial networks~\cite{LSRV20}, and chemical networks~\cite{KMBPR16}.
Because of their non-Euclidean nature, graphs do not present familiar features that are common to other systems, like a coordinate or vector space, making the analysis of graph data challenging.
To address this issue, the graph embedding algorithms have been proposed to obtain effective graph data representation that represents graphs concisely in Euclidean space~\cite{PAS14,TQWZYM15,GL16}.
The core idea of those algorithms is to transform graphs from non-Euclidean space into low dimensional vectors, in which the graph information is implicitly preserved. 
After the transformation, a plethora of downstream tasks can be efficiently performed, such as node classification~\cite{GL16,DBV16} and graph classification~\cite{YYMRHL18}.

Recently, a new family of deep learning models known as graph neural networks (GNNs) has been proposed to obtain the graph embedding and achieved state-of-the-art performance.
The core idea of GNNs is to train a deep neural network that aggregates the feature information from neighborhood nodes to obtain \textit{node embedding}.
They can be further aggregated to obtain the \textit{graph embedding} for graph classification.
Such graph embedding is empirically considered sanitized since the whole graph is compressed to a single vector. 
In turn, it has been shared with third parties to conduct downstream graph analysis tasks.
For example, the graph data owner can generate the graph embeddings locally and upload them to the Embedding Projector service\footnote{\url{https://projector.tensorflow.org/}} provided by Google to visually explore the properties of the graph embeddings.
Despite that sharing graph embeddings for downstream graph analysis tasks is intriguing and practical, the associated security and privacy implications remain unanswered.

\mypara{Our Contributions}
In this paper, we initiate a systematic investigation of the privacy issue of graph embedding by exploring three inference attacks.
The first attack is \textit{property inference} attack, which aims to infer the basic properties of the target graph given the graph embedding, such as the number of nodes, the number of edges, the graph density, etc.
We then investigate the \textit{subgraph inference} attack.
That is, given the graph embedding and a subgraph of interest, the adversary aims to determine whether the subgraph is contained in the target graph.
For instance, an adversary can infer whether a specific chemical compound structure is contained in a molecular graph if gaining access to its graph embedding, posing as a direct threat to the intellectual property of the data owner.
The challenge of subgraph inference attack is that the formats of the graph embedding (\ie a vector) and the subgraph of interest (\ie a graph) are different and not directly comparable.
Finally, we aim to reconstruct a graph that shares similar structural properties (\eg degree distribution, local clustering coefficient, etc.) with the target graph.
We call this attack \textit{graph reconstruction} attack.
For instance, if the target graph is a social network, the reconstructed graph would then allow an adversary to gain direct knowledge of sensitive social relationships. 
In summary, we make the following contributions.
\begin{itemize}[leftmargin=*]
    \setlength\itemsep{-0.25em}
    \item To launch the property inference attack, we model the attack as a multi-task classification problem, where the attack model can predict all the graph properties of interest simultaneously.
    We conduct experiments on five real-world graph datasets and three state-of-the-art graph embedding models to validate the effectiveness of our proposed attack.
    The experimental results show that we can achieve up to 0.89 attack accuracy on the DD dataset.
    \item We design a novel graph embedding extractor, enabling the subgraph inference attack model to simultaneously learn from both the graph embedding and the subgraph of interest.
    The experimental results on five datasets and three graph embedding models validate the effectiveness of our attack.
    For instance, we achieve 0.98 attack AUC on the DD dataset.
    We further successfully launch two transfer attacks when the sampling method and embedding model architecture for training and testing attack model are different.
    \item We propose to use the graph auto-encoder paradigm to mount the graph reconstruction attack.
    Once the graph auto-encoder is trained, its decoder is employed as our attack model.
    Extensive experiments show that the proposed attack can achieve high similarity in terms of graph isomorphism and macro-level graph statistics such as degree distribution and local clustering coefficient distribution.
    For instance, the cosine similarity of local clustering coefficient distribution between the target graph and the reconstructed graph can achieve 0.99.
    The results exemplify the effectiveness of our graph reconstruction attack.
    \item To mitigate the inference attacks, we further propose a defense mechanism based on graph embedding perturbation.
    The main idea is to add well-calibrated Laplace noise to the graph embedding before sharing with third parties.
    We demonstrate through several experiments that our proposed defense can effectively mitigate all the three inference attacks without noticeable performance degradation for graph classification tasks.
\end{itemize}

\section{Preliminaries}
\label{sec:background}

\subsection{Notations}
\label{sec:notations}
We denote an undirected, unweighted, and attributed graph by $\graph = \tuple{\nodeset, \adj, \feat}$, where $\nodeset$ represents the set of all nodes, $\adj$ is the adjacency matrix, $\feat$ is the attributes matrix.
We denote the embedding of a node $u \in \nodeset$ as $\embed_u$ and the whole graph embedding as $\embed_\graph$ (see~\autoref{sec:gnn_intro} for details).
We summarize the frequently used notations introduced here and in the following sections in~\autoref{table:notations} of \autoref{app:notations}.

\subsection{Graph Neural Network}
\label{sec:gnn_intro}
Many important real-world datasets are in the form of graphs, e.g., social networks~\cite{QTMDWT18}, financial networks~\cite{LSRV20}, and chemical networks~\cite{KMBPR16}. 
The classical machine learning architectures and algorithms oftentimes do not perform well with these kinds of data. 
Most of them were designed to learning from data that can naturally be represented individually (\ie data points) but are less effective in dealing with relational data with more complex structure.
To effectively extract useful information from the graph data, a new family of deep learning algorithms, i.e., graph neural networks (GNNs), has been proposed and achieved superior performance in various tasks~\cite{AT16,DBV16,KW17,VCCRLB18}. 
GNNs generalize the deep neural network models to graph-structured data and learn representations for graph-structured data by aggregating information from a node’s neighbors using neural networks, i.e., learning a model $\model: \graph \rightarrow \embed$. The learned embedding \embed can be used for different graph analytics tasks - node classification~\cite{HYL17, KW17}
and graph classification~\cite{YYMRHL18, XHLJ19}.

\begin{itemize}[leftmargin=*]
    \setlength\itemsep{-0.25em}
    \item \textbf{Node Classification.} 
    The objective of node classification is to determine the label of nodes in the graph, such as the gender of a user in a social network. 
    GNNs first generate node embeddings $\embed_u$, and feed them to a classifier to determine the node labels.
    \item \textbf{Graph Classification.} 
    The objective of graph classification is to determine the label of the whole graph, such as a molecule's solubility or toxicity. 
    In graph classification, one needs to further transform all the node embeddings $\embed_u, \forall u \in \nodeset$ to a whole graph embedding $\embed_\graph$ to determine the label of the whole graph.
\end{itemize}

\subsubsection{Message Passing}
Most of the existing GNNs use \textit{message passing} to obtain the \textit{node embedding} $\embed_u$.
It starts by assigning the node attributes as the node embeddings.
Then, every node receives a ``message'' from its neighbor nodes and aggregates the messages as its intermediate embedding.
After $K$ steps, the node embedding aggregates information from its $K$-hop neighbors.
Formally, during each message passing iteration, the node embedding $\embed_u^{k}$ of node $u \in \nodeset$ is updated using ``message'' aggregated from $u$'s graph neighborhood $\neigh{u}$ using a pair of \textit{aggregation operation} $\aggr$ and \textit{updating operation} $\upd$:
\begin{align*}
    \embed_u^{k+1} 
    = \update{k}{\embed_u^{k}, \messagee_{\neigh{u}}^k}
    = \update{k}{\embed_u^{k}, \aggregate{k}{\embed_v^{k}, \forall v \in \neigh{u}}}
\end{align*}
where $\embed_u^k \in \mathbb{R}^{n \times d_\embed}$ is the node embedding of node $u$ after $k$ steps of message passing, $\messagee_{\neigh{u}}^k$ is the message received from node $u$'s neighborhood $\neigh{u}$, which is calculated by $\aggr$.

\mypara{Aggregation Operation}
Recently, researchers have proposed many practical implementations of \aggr.
Graph Isomorphism Networks (\gin)~\cite{XHLJ19} uses \emph{sum} operation to aggregate the embeddings of all node $u \in \graph_{\neigh{u}}$.
Graph SAmple and aggreGatE (\sage)~\cite{HYL17} uses \emph{mean} operation to aggregate all node embeddings of $\graph_{\neigh{u}}$ instead of summing them up.
The Graph Convolution Networks (\gcn)~\cite{KW17} method uses the symmetric normalization, and the Graph Attention Networks (\gat)~\cite{VCCRLB18} method uses the attention mechanism to learn a weight matrix to aggregate the embeddings of all node $u \in \graph_{\neigh{u}}$.

\mypara{Updating Operation}
The updating operation $\upd$ combines the node embeddings from node $u$ and the message from $u$'s neighborhood.
The most straightforward updating operation is to calculate the weighted combination~\cite{SGTHM09}.
Formally, we denote the basic updating operation as $\upd_{base} = \sigma(W_{self}\embed_u + W_{neigh}\messagee_{\neigh{u}})$, where $W_{self}$ and $W_{neigh}$ are learnable parameters, $\sigma$ is a non-linear activation function.
Another method is to treat the basic updating operation as a building block, and concatenate it with the current embedding~\cite{HYL17}.
We denote the concatenation-based updating operation as $\upd_{concat} = \upd_{base} || \embed_u$, where $||$ is the concatenation operation.
An alternative is to use the weighted average of the basic updating method and the current embedding~\cite{PTPV17}, which is referred as \textit{interpolation-based} updating operation and is formally defined as $\upd_{inter} = \alpha_1 \circ \upd_{base} + \alpha_2 \circ \embed_u$.

\subsubsection{Graph Pooling}
The \textit{graph pooling operation} \pool aggregates the embeddings of all nodes in the graph to form a whole graph embedding, i.e., $\embed_\graph = \pooling{\embed_u, \forall u \in \graph}$.

\mypara{Global Pooling}
The most straightforward approach for graph pooling is to directly aggregate all the node embeddings, which is called global pooling, such as \textit{max pooling} and \textit{mean pooling}.
Although simple and efficient, the global pooling operation could lose the graph structural information, leading to unsatisfactory performance~\cite{YYMRHL18,BGA20}.

\mypara{Hierarchical Pooling}
To better capture the graph structural information, researchers have proposed many hierarchical pooling methods~\cite{YYMRHL18,BGA20}.
The general idea is to aggregate $n$ node embeddings to one graph embedding hierarchically, instead of aggregating them in one step as global pooling.
Concretely, we first obtain $n$ node embeddings using message passing modules, and finds $m$ clusters according to the node embeddings, where $1<m<n$.
Next, we treat each cluster as a node with features being the graph embedding of this cluster, then iteratively applying the message passing and clustering operations until there are only one graph embedding.

Formally, in the $\ell$-th pooling step, we need to learn a cluster assignment matrix $\assign^\layer \in \mathbb{R}^{n_\ell \times n_{\layer + 1}}$, which provides a soft assignment of each node at layer $\layer$ to a cluster in the next coarsened layer $\layer + 1$.
Suppose $S^\layer$ in layer $\layer$ has already been computed, we can use the following equations to compute the coarsened adjacency matrix $\adj^{\layer + 1}$ and a new matrix of node embeddings $\embed^{\layer + 1}$:
\begin{align*}
    \embed^{\layer + 1} &= {\assign^\layer}^T \embed^{\layer} \in \mathbb{R}^{n_{\layer + 1} \times d_\embed} \\
    \adj^{\layer + 1} &= {\assign^\layer}^T \adj^{\layer} \assign^{\layer} \in \mathbb{R}^{n_{\layer + 1} \times n_{\layer + 1}}
\end{align*}
The main challenge lies in how to learn the cluster assignment matrix $\assign^\layer$.
In the following, we introduce two state-of-the-art methods.

\begin{itemize}[leftmargin=*]
    \setlength\itemsep{-0.25em}
    \item \mypara{Differential Pooling~\cite{YYMRHL18}}
    The \diffpool method uses a message passing module to calculate the assignment matrix as
    $
        \assign^{\layer} = \mbox{softmax}\left( \mbox{GNN}(\adj^{\layer}, \embed^{\layer}) \right)
    $.
    In practice, it can be difficult to train the GNN models using only gradient signal from the output layer.
    To alleviate this issue, \diffpool introduces an auxiliary link prediction objective to each pooling layer, which encodes the intuition that nearby nodes should be pooled together.
    In addition, \diffpool introduces another objective to each pooling layer that minimizes the entropy of the cluster assignment.
    
    \item \mypara{MinCut Pooling~\cite{BGA20}}
    The \mincutpool method uses an MLP (multi-layer perceptron) module to compute the assignment matrix as
    $
        \assign^\layer = \mbox{softmax}\left( \mbox{MLP}(\adj^\layer, \embed^\layer) \right)
    $.
    Different from \diffpool, \mincutpool introduces the minimum cut objective to each pooling layer that aims to remove the minimum volume of edges, which is in line with the objective of graph pooling aiming to assign the closely connected nodes into the same cluster.
\end{itemize}

\mypara{Implementation of GNN Model}
Typically, the \textit{graph-level} GNN models consist of a graph embedding module, which encode the graph into the graph embedding, and a multi-class classifier, which predict the label of the graph using the graph embedding.
To train the GNN model, we normally adopt the cross-entropy loss.
For graph embedding modules containing hierarchical pooling operations, we need to incorporate additional loss such as minimum cut loss in \mincutpool.
After the GNN model is trained, we use the graph embedding module as our embedding generation model in the following parts.

\begin{figure*}[!th]
\begin{center}
\includegraphics[width=0.8\textwidth]{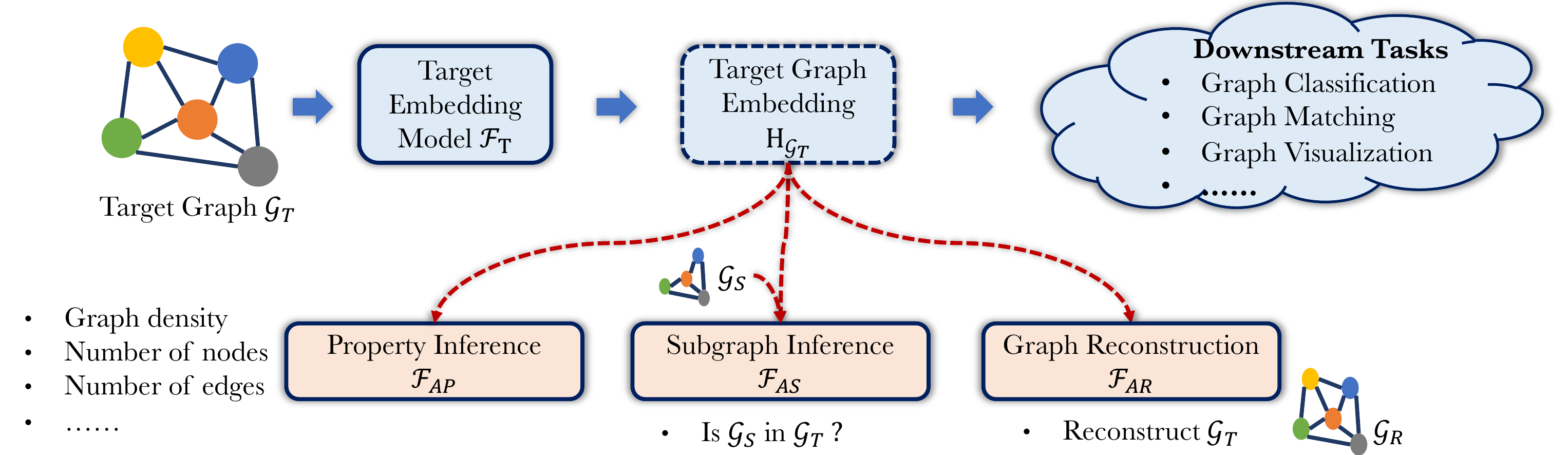}
\end{center}
\vspace{-0.3cm}
\caption{
Attack taxonomy of the graph embedding.
The adversary obtains the whole graph embedding $\embed_{\targetgraph}$ of a sensitive target graph \targetgraph, which is primarily shared to third parties for downstream tasks, and aims to infer sensitive information about \targetgraph: (1) Infer the basic properties of \targetgraph, such as the number of nodes, the number of edges, and graph density (\propinfermodel); 
(2) given a subgraph of interest \subgraph, infer whether \subgraph is contained in \targetgraph (\subinfermodel); 
(3) reconstruct a graph \recongraph that is similar with \targetgraph (\graphreconmodel).
}
\label{fig:threat_model}
\end{figure*}

\section{Threat Model and Attack Taxonomy}
\label{sec:threat_model}

\subsection{Motivation} 
In this paper, we focus on the whole graph embedding $\embed_\graph$, which is oftentimes computed on a sensitive graph (\eg biomedical molecular network and social network).
Such graph embedding $\embed_\graph$ is empirically considered sanitized since the whole graph is compressed to a single vector. 
In practice, it has been shared with third parties to conduct downstream graph analysis tasks.
For example, the graph data owner can calculate the graph embeddings locally and upload them to the Embedding Projector service provided by Google to visually explore the properties of the graph embeddings.
Another example is that some companies release their graph embedding systems, together with which they publish some pretrained graph embeddings to facilitate the downstream tasks.
These systems including the PyTorch BigGraph\footnote{\url{https://github.com/facebookresearch/PyTorch-BigGraph}} system developed by Facebook, DGL-KE\footnote{\url{https://github.com/awslabs/dgl-ke}} system developed by Amazon, and GROVER developed by Tencent\footnote{\url{https://github.com/tencent-ailab/grover}}.
Besides, the graph embeddings can also be shared in the well-known model partitioning paradigm~\cite{LG15, KHGRMMT17}.
This paradigm can effectively improve the scalability of inference by allowing the graph data owner to calculate the graph embeddings locally, and upload them to the cloud for further inference or analysis.

Despite sharing graph embeddings for downstream graph analysis tasks is intriguing and promising, the associated security and privacy implications remain unanswered. 
For instance, Song~\etal~\cite{SS20,SR20} demonstrated that the embeddings can leak sensitive information about image and text data in the Euclidean space.
Recall that the goal of graph embedding $\embed_\graph$ is to preserve graph-level similarity, a natural question is: 
would the graph embedding $\embed_\graph$ leak sensitive structural information of its corresponding graph $\graph$?

\subsection{Threat Model}
We consider the scenario where the adversary obtains a whole graph embedding (which is referred to as \textit{target graph embedding} $\embed_{\graph_T}$) from the victim, either from Embedding Projector, pretrained graph embeddings, or model partitioning paradigm.
The goal of the adversary is to infer the sensitive information of the graph that is used to generate this graph embedding.
We call this graph \textit{target graph} $\graph_T$, and the GNN model that used to generate the target graph embedding \textit{\targetmodelname} $\targetmodel$.
Note that inferring the sensitive information of target graph with ``graph embedding'' is more challenging than that with ``node embeddings'' in previous study~\cite{DBS20}.
From the attacker's perspective, it represents the most difficult setting since the whole graph is compressed to a single vector by the aforementioned pooling methods in Section~\ref{sec:background}.
To train the attack model $\attackmodel$, we assume the adversary has an auxiliary dataset $\daux$ that comes from the same distribution of the target graph.
This is plausible in practice. 
For instance, if the target graph embedding is generated from a social network, the adversary can collect social network graphs by themselves through public data API.\footnote{\url{https://developer.twitter.com/en/docs/twitter-api}}
For molecular networks, the adversary can use the public datasets online.\footnote{\url{https://chrsmrrs.github.io/datasets}}
We also show that our attacks is still effective when \daux comes from different distribution than the target graphs in \autoref{sec:experiment}.
We further assume the adversary only has black-box access to the \targetmodelname~\cite{SS20,SR20}, which is the most difficult setting for the adversary~\cite{SSSS17,JCBKP20,SRS17,MSCS19,SRS17}.
This assumption is plausible when the \targetmodelname is accessible via public API or freely available online.\footnote{\url{http://snap.stanford.edu/gnn-pretrain/}}

\subsection{Attack Taxonomy}
We formalize three inference attacks that can reveal sensitive information of the target graph given the threat model.
An overview of the attack taxonomy is shown in \autoref{fig:threat_model}.

\mypara{Property Inference Attack ($\propinfermodel$)}
Given the target graph embedding \targetembed, the attack goal is to infer the basic properties of \targetgraph, such as the number of nodes, the number of edges, the density, etc.

Note that the primary goal of GNN is learning information from graphs for downstream tasks, e.g., protein toxicity prediction. 
Many graph properties, such as node numbers, are not related to the downstream tasks, and successful property inference attacks imply such properties are overlearned~\cite{SS20,SR20} by GNNs. 
These properties can be proprietary when the graph contains valuable information such as molecules. 
Inferring such properties can directly violate the intellectual property (IP) of the data owner.

\mypara{Subgraph Inference Attack ($\subinfermodel$)}
Given the target graph embedding \targetembed and a subgraph of interest \subgraph, the attack goal is to infer whether \subgraph is contained in \targetgraph. 
For instance, an attacker can infer whether a specific chemical compound structure (\subgraph) is contained in a molecular graph ($\graph_T$) if gaining access to its graph embedding ($\embed_{\graph_T}$).
Note that we consider the scenario where the subgraph constituting a major part of the target graph.
Small graphs, such as triangles or stars, are universal for almost all graphs, hence not taking part in our subgraph inference attack.

\mypara{Graph Reconstruction Attack ($\graphreconmodel$)}
Given the graph embedding \targetembed, the attack goal is to reconstruct a graph \recongraph that shares similar graph structural statistics, such as degree distribution and local clustering coefficient, with \targetgraph.
Concretely, we aim to reconstruct an adjacency matrix \adj of \targetgraph.
Knowing the high-level structural quantities of the molecular graphs may lead to IP loss of the companies creating them.
For instance, the adversary can develop generic drugs with much lower cost than the famous pharmaceutical companies by exploiting the high-level structural quantities of the reconstructed molecular graphs to narrow down the search space.

\section{Property Inference Attack}
\label{sec:property_infer}

\subsection{Attack Overview}
\begin{figure}[!ht]
\begin{center}
\includegraphics[width=0.4\textwidth]{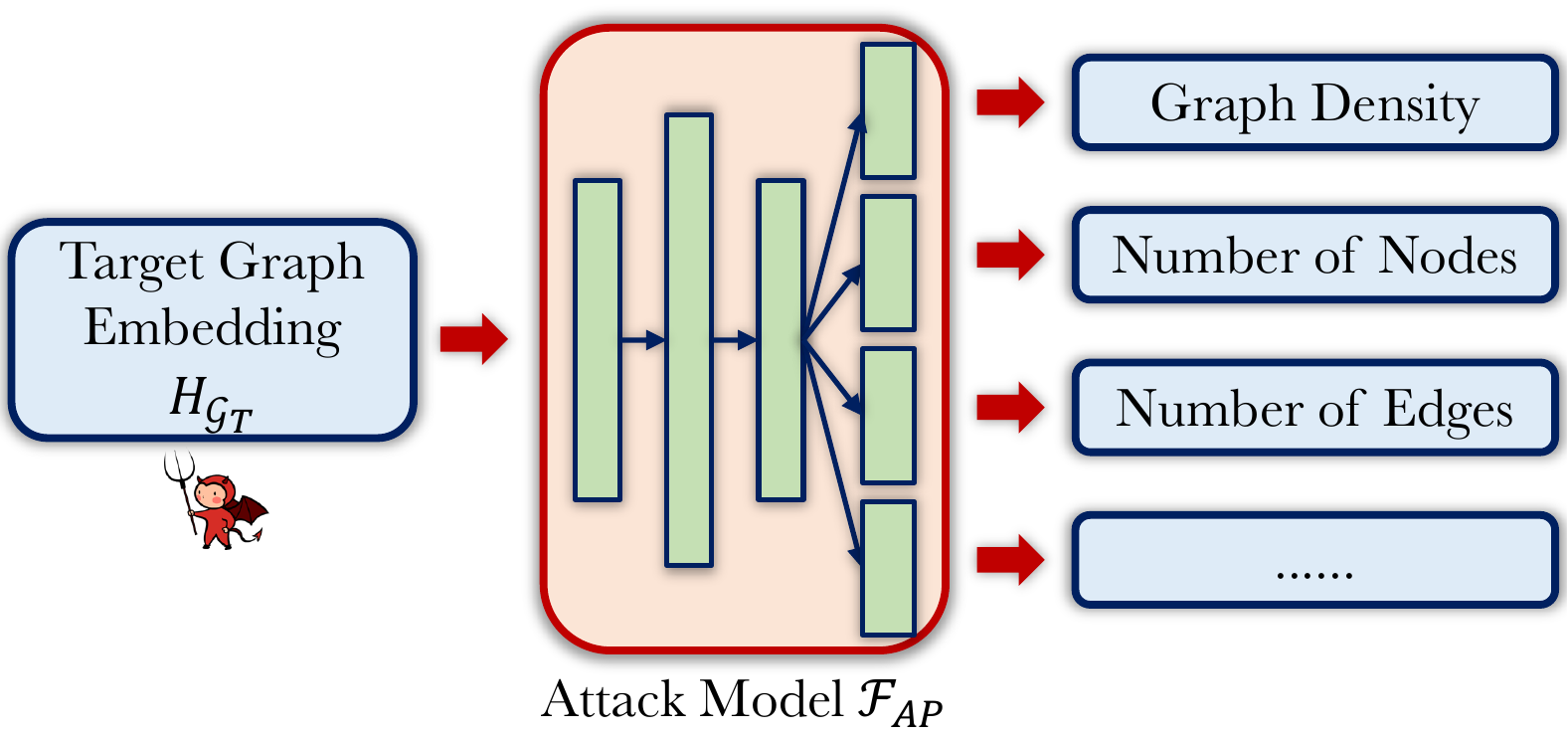}
\end{center}
\vspace{-0.3cm}
\caption{Attack pipeline of the property inference attack. 
The attack model \propinfermodel is a multi-task classifier, which consists of multiple output layers, each predicts one graph property.}
\label{fig:property_infer}
\end{figure}

Given the target graph embedding \targetembed, the goal of the property inference attack is to infer the basic properties of the target graph \targetgraph, such as the number of nodes, the number of edges, and density.
\autoref{fig:property_infer} illustrates the general attack pipeline of the property inference attack. 
Our attack model \propinfermodel takes as input the target graph embedding \targetembed and outputs all the interested graph properties of \targetgraph simultaneously.

\subsection{Attack Model $\propinfermodel$}

\mypara{Model Definition}
Formally, the property inference attack $\propinfermodel$ is defined as
\begin{align*}
    \propinfermodel: \targetembed \rightarrow \{\mbox{graph properties}\}
\end{align*}
\noindent Concretely, the attack model consists of a feature extractor $\mathcal{E}$ (multiple sequential linear layers), and multiple parallel prediction layers $\mathcal{M}$, each is responsible for predicting one property. 
We outline the technical details of building $\propinfermodel$ below.

\mypara{Training Data}
To train the attack model, we need a set of graph embeddings $\embed_\graph$ and a set of properties of interest \property.
As discussed in~\autoref{sec:threat_model}, the adversaries have access to an auxiliary dataset \daux that comes from the same distribution of \targetgraph.
The adversaries can obtain the auxiliary graph embedding \auxembed of the auxiliary graph $\auxgraph \in \daux$ by querying the \targetmodelname.
Finally, we use the graph properties of \auxgraph to label \auxembed. 
We further bucketize the domain of the property values into $k$ bins.
For instance, the density of a graph is in the range of $[0, 1]$ and $k=5$, we bucketize the graph density into $5$ bins, which results in $5$ classes in the classification.
Note that modeling the inference of continuous value into multi-class classification is commonly used, such as demographic properties prediction in social networks~\cite{MW16} and dropout rate prediction~\cite{OASF18}.

\mypara{Training Attack Model}
Recall that the attack model \propinfermodel is the combination of a feature extractor $\mathcal{E}$ and multiple prediction layers $\mathcal{M}$, we can train the attack model by optimizing the following optimization problem:
\begin{align*}
    \min \; \underset{\auxgraph \in \daux}{\mathbb{E}}\left[\sum_{p \in \mathbb{P}} \loss\left[ \mathcal{M}^p(\mathcal{E}(\auxembed)), p \right] \right]
\end{align*}
where $\mathbb{P}$ is the set of properties that the attackers interested, $p$ is a property in $\mathbb{P}$, \loss is the cross-entropy loss.
Notice that all properties share the same parameters for $\mathcal{E}$, and use different parameters for $\mathcal{M}^p$.

\section{Subgraph Inference Attack}
\label{sec:subgraph_infer}

\subsection{Attack Overview}

\begin{figure}[!ht]
\begin{center}
\includegraphics[width=0.45\textwidth]{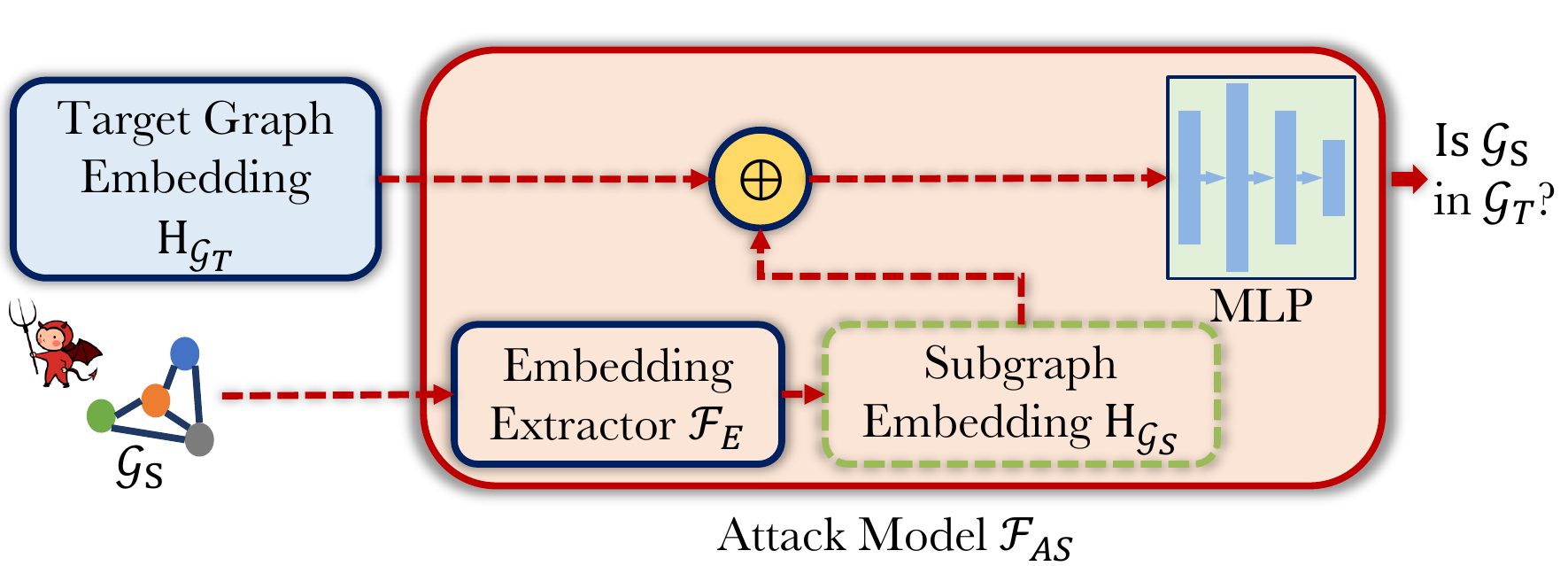}
\end{center}
\vspace{-0.3cm}
\caption{Attack pipeline of the subgraph inference attack.
The attack model \subinfermodel has two inputs with different formats, namely target graph embedding and subgraph.
The subgraph is transformed to a subgraph embedding by an embedding extractor integrated in the attack model, aggregated with the target embedding, and sent to a binary classifier for prediction.
}
\label{fig:subgraph_infer}
\end{figure}

Given the target graph embedding \targetembed and a subgraph of interest \subgraph, the attack goal is to infer whether \subgraph is contained in \targetgraph. 
Here, we assume that \subgraph constitutes a major part of the target graph \targetgraph.\footnote{We experiment with subgraphs containing from 20\% to 80\% of the target graph's nodes (see \autoref{subsec:evaluation_subinfer}).}
That is, we do not focus on small subgraphs, such as triangles or stars, as they appear in almost all the graphs, thus not worth the adversary's efforts.
The general attack pipeline of subgraph inference attack is illustrated in~\autoref{fig:subgraph_infer}.

Note that subgraph inference attack is more challenging than the property inference attack $\propinfermodel$. 
First, subgraph isomorphism is known to be NP-complete~\cite{GJ79}.
Second, the attack model \subinfermodel has two inputs with different formats, namely the embedding (\targetembed) and the graph (\subgraph), and cannot be directly compared.
To make the two inputs comparable, we integrate a \textit{graph embedding extractor} $\model_E$ in the attack model to transform the subgraph \subgraph to a subgraph embedding \subembed.
The architecture of $\model_E$ can be either the same with (when the \targetmodelname is known) or different from (when the \targetmodelname is unknown) the \targetmodelname \targetmodel.
Finally, the target graph embedding \targetembed and the subgraph embedding \subembed are aggregated, using the approaches introduced in~\autoref{subsec:subgraph_model_train}, and sent to a binary classifier for prediction.

\subsection{Attack Model $\subinfermodel$}
\label{subsec:subgraph_model_train}

\mypara{Attack Definition}
Formally, the subgraph inference attack is defined as
\begin{align*}
    \subinfermodel: \tuple{\targetembed, \subgraph} \rightarrow \{\subgraph \in \targetgraph, \subgraph \notin \targetgraph\}
\end{align*}

\noindent Concretely, the attack model $\subinfermodel$ is a binary classifier to determine if a given subgraph $\subgraph$ is contained in the target graph $\targetgraph$.
We outline the technical details of building $\subinfermodel$ below.

\mypara{Generating Positive and Negative Samples}
Similar to the property inference attack, we use the auxiliary dataset \daux to obtain the training data for the attack model \subinfermodel.
To generate ground truth for \subinfermodel, given an auxiliary graph $\auxgraph \in \daux$, we generate a \textit{positive subgraph} $\subgraph \in \auxgraph$ and a \textit{negative subgraph} $\Bar{\subgraph} \notin \auxgraph$.
The positive subgraph \subgraph is generated by sampling a subgraph from the auxiliary graph \auxgraph using the graph sampling method, such as \textit{random walk}.
To generate the negative subgraph $\Bar{\subgraph}$, we use the same sampling method to sample a subgraph from another auxiliary graph $\auxgraph' \in \daux$ and $\auxgraph' \neq \auxgraph$.
As aforementioned, the subgraph of interest constitutes a major part of the target graph, the sampled negative subgraph $\Bar{\subgraph}$ is unlikely to be contained in \auxgraph.

For each auxiliary graph \auxgraph, we have one positive subgraph \subgraph and one negative subgraph $\Bar{\subgraph}$.
The adversary first obtains the auxiliary graph embedding \auxembed by querying the \targetmodelname.
They then have a \textit{positive sample} $\tuple{\auxembed, \subgraph}$, which is labeled as $1$, and a negative sample $\tuple{\auxembed, \Bar{\subgraph}}$, which is labeled as $0$, for the attack model.

\mypara{Constructing Features}
The attack model first uses a graph embedding extractor to transform the subgraph \subgraph into a subgraph embedding \subembed to make the two inputs comparable.
The attack model then aggregates the target graph embedding \targetembed and the subgraph embedding \subembed to generate an \textit{attack feature vector} \subfeat.
In this paper, we propose the following three aggregation strategies:

\begin{itemize}[leftmargin=*]
    \setlength\itemsep{-0.25em}
    \item \mypara{Concatenation}
    A commonly used approach is to concatenate the two graph embeddings, i.e., $\subfeat = \targetembed || \subembed$, where $||$ is the concatenation operation.
    \item \mypara{Element-wise Difference}
    An alternative is to calculate the element-wise difference of two graph embeddings, i.e.,
    $\subfeat = \targetembed - \subembed$.
    \item \mypara{Euclidean Distance}
    Another approach is to calculate the Euclidean distance between two graph embeddings, i.e., $\subfeat = ||\targetembed - \subembed||_2$.
\end{itemize}

We empirically evaluate the effectiveness of these three strategies in \autoref{subsec:evaluation_subinfer}.

\mypara{Training Attack Model}
The final step of the attack is to send the attack feature vector \subfeat to a binary classifier, which is modeled as an MLP (multi-layer perceptron), to determine whether \subgraph is contained in \targetgraph.
We use the cross entropy loss and gradient decent algorithm to train the attack model.
Note that the binary classifier and the graph embedding extractor in the attack model \subinfermodel are trained simultaneously.

\section{Graph Reconstruction Attack}
\label{sec:graph_reconstruct}

\subsection{Attack Overview}
\begin{figure}[!ht]
\begin{center}
\includegraphics[width=0.45\textwidth]{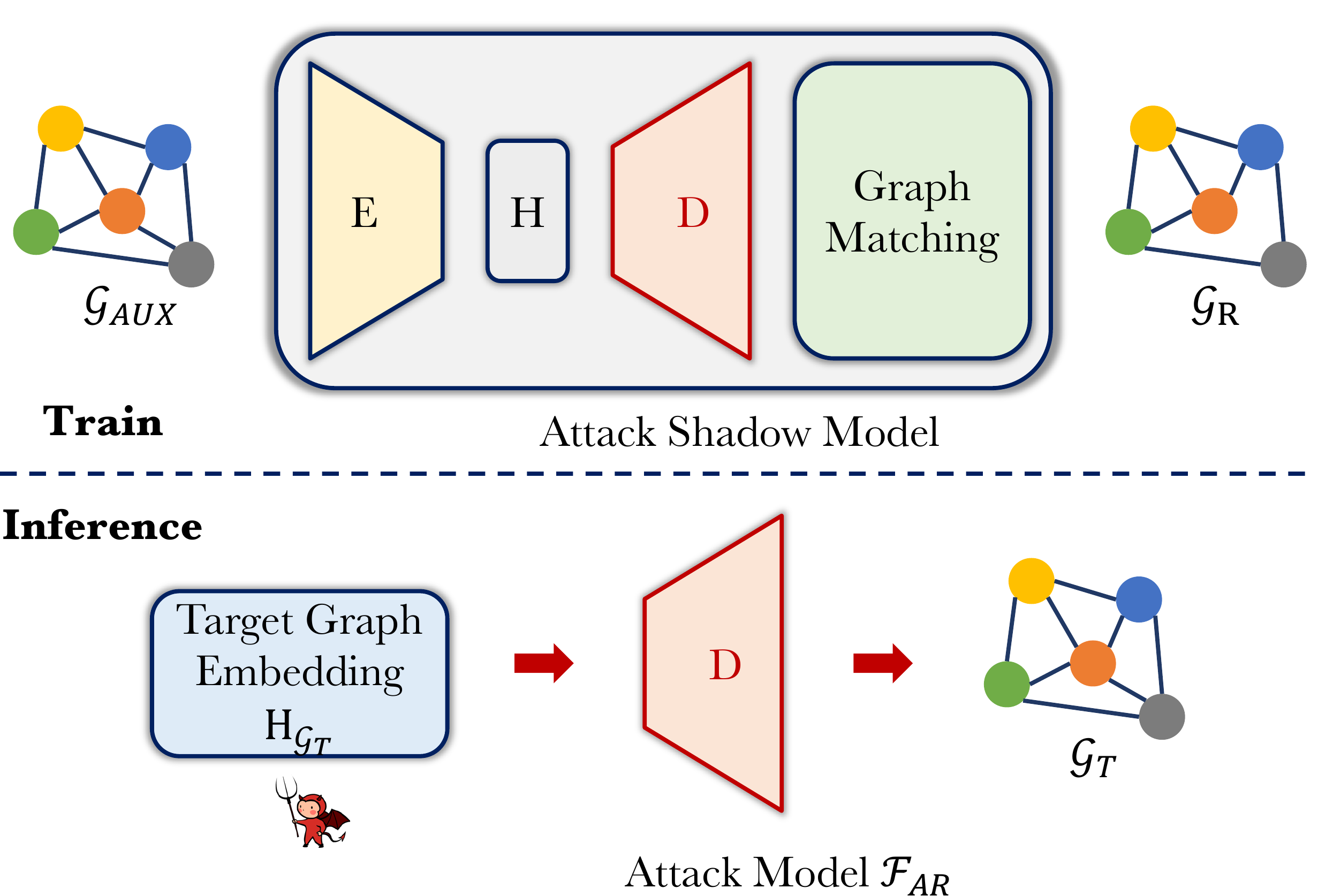}
\end{center}
\vspace{-0.3cm}
\caption{Attack pipeline of the graph reconstruction attack.
The attack model \graphreconmodel is a decoder that can transform the graph embedding to a graph.
The decoder can be obtained from the graph auto-encoder paradigm.}
\label{fig:graph_reconstruct}
\end{figure}

Given the target graph embedding \targetembed, the attack goal is to reconstruct a graph \recongraph that have similar graph statistics, such as degree distribution and local clustering coefficient, with the target graph \targetgraph.
\autoref{fig:graph_reconstruct} shows the overall attack pipeline of graph reconstruction attack.
The graph reconstruction attack is the most challenging task because we are rebuilding the whole graph from a single vector $\embed_G$. 
To this end, the attack model \graphreconmodel leverages a tailored graph auto-encoder~\cite{SK18} and puts its decoder into service to transform the graph embedding to a graph.
Once trained, the adversary feeds \targetembed to the decoder, and the decoder would output reconstructed graph \recongraph that have similar graph statistics with the target graph \targetgraph.

\subsection{Attack Model $\graphreconmodel$}
\label{subsec:gae_train}

\mypara{Attack Definition}
Formally, the graph reconstruction attack is defined as
\begin{align*}
    \graphreconmodel: \targetembed \rightarrow \recongraph
\end{align*}

\noindent Essentially, the graph reconstruction attack $\graphreconmodel$ is the decoder of a customized graph auto-encoder.
We outline the technical details of building $\graphreconmodel$ below.

\mypara{Graph Auto-encoder Design}
We use the graph auto-encoder paradigm to train the attack model. 
The architecture is shown in the training phase of~\autoref{fig:graph_reconstruct}.
We use an auxiliary dataset \daux to train the graph auto-encoder.
Different from the auto-encoder in the image domain, the graph auto-encoder has an additional component named \textit{graph matching} except for the encoder and decoder.
The reason for introducing the graph matching component is that neither the auxiliary graph $\auxgraph \in \daux$ nor the reconstructed graph \recongraph imposes node orderings (i.e., graph isomorphism), making the calculation of loss between \auxgraph and \recongraph inaccurate.
For instance, an auxiliary graph \auxgraph and a reconstructed graph \recongraph with the same structure and completely different node orderings can have different adjacency matrix, such that the loss between \auxgraph and \recongraph is large while it is expected to be zero.
Besides, the encoder in the graph auto-encoder can transform a graph to the graph embedding, which can be modeled as a GNN model.
The decoder can transform the graph embedding back to graph in the form of an adjacency matrix, which can be modeled as a multi-layer perceptron.

\mypara{Graph Matching}
Following the same strategy as in~\cite{SK18}, we adopt the maximum pooling matching method in our implementation.
The main idea is to find a transformation matrix $Y \in \{0, 1\}^{n \times n}$ between \targetgraph and \recongraph, where $Y_{a,i} = 1$ if node $v_a \in \targetgraph$ is assigned to $v_i \in \recongraph$, and $Y_{a,i} = 0$ otherwise.
Due to space limitation, we refer the readers to~\cite{SK18} for the detailed calculation of $Y$. 

\mypara{Training Attack Model}
To train the graph auto-encoder, we use the cross entropy to calculate the loss between \auxgraph and \recongraph, which calculates the cross entropy between each pair of elements in \auxgraph and \recongraph.
Formally, denote the adjacency matrix of \auxgraph and \recongraph as $\adj_{\auxgraph}$ and $\adj_{\recongraph}$ respectively.
For each training sample, we first conduct the graph matching to obtain $Y$, then use the cross entropy between $\adj_{\auxgraph}$ and $Y\adj_{\recongraph} Y^T$ to update the graph auto-encoder.

\mypara{Fine-tuning Decoder}
Note that the structure or the parameters of the encoder can be different from the \targetmodelname; thus, the decoder may not perfectly capture the correlation between the auxiliary graph \auxgraph and its graph embedding \auxembed generated by the \targetmodelname.
To address this issue, we use the auxiliary graph \auxgraph to query the \targetmodelname and obtain the corresponding graph embedding $\embed_{\auxgraph}$.
Then, the graph-embedding pairs $\tuple{\auxgraph,\embed_{\auxgraph}}$ obtained from the \targetmodelname are used to fine-tune the decoder using the same procedure of graph matching and loss function as aforementioned~\cite{SBBFZ20}.

\mypara{Discussion}
Both the space and time complexity of the graph matching algorithm are $O(n^4)$; thus, our attack can be only applied to graphs with tens of nodes.
This is enough in many real-world datasets, such as bioinformatics graphs and mocecular graphs.
In the future, we plan to investigate more advanced methods to extend our attacks to larger graphs.
Besides, our current attack can only restore the graph structure of the target graph.
We plan to reconstruct the node features and the graph structure simultaneously in the future.

\section{Evaluation}
\label{sec:experiment}

\subsection{Experimental Setup}

\mypara{Datasets}
We conduct our experiments on five public graph datasets from TUDataset~\cite{MKBKMN20}, including DD, ENZYMES, AIDS, NCI1, and OVCAR-8H.
These datasets are widely used as benchmark datasets for evaluating the performance of GNN models~\cite{XHLJ19,CVCB19,EPBM20,DJLBB20}. 
DD and ENZYMES are bioinformatics graphs, where the nodes represent the secondary structure elements, and an edge connects two nodes if they are neighbors along the amino acid sequence or one of three nearest neighbors in space.
The node features consist of the amino acid type, i.e., helix, sheet, or turn, as well as several physical and chemical information.
AIDS, NCI1, and OVCAR-8H are molecule graphs, where nodes and edges represent atoms and chemical bonds, respectively.
The node features typically consist of one-hot encoding of the atom type, e.g., hydrogen, oxygen, carbon, etc.
Each dataset has multiple independent graphs with a different number of nodes and edges, and each graph is associated with a label.
For instance, the label of the molecule datasets indicates the toxicity or biological activity determined in drug discovery projects.
\autoref{table:dataset_statistic} summarizes the statistics of all the datasets.

\begin{table}
\centering
\caption{Dataset statistics, including the type of graphs, the total number of graphs in the dataset, the average number of nodes, the average number of edges, and the number of classes associated with each dataset.
The datasets with \textsuperscript{$\ast$} are used for dataset transfer attacks.}
\label{table:dataset_statistic}
\scriptsize
\setlength{\tabcolsep}{0.25em}
\renewcommand{\arraystretch}{1.0}
\begin{tabular}{c | c c c c c c}
\toprule
Dataset & Type & \# Graphs & Avg. Nodes & Avg. Edges & \# Feats & \# Classes \\
\toprule
DD & Bioinformatics & 1,178 & 284.32 & 715.66 & 89 & 2 \\
ENZYMES & Bioinformatics & 600 & 32.63 & 62.14 & 21 & 6 \\
AIDS & Molecules & 2,000 & 15.69 & 16.20 & 42 & 2 \\
NCI1 & Molecules & 4110 & 29.87 & 32.30 & 37 & 2 \\
OVCAR-8H & Molecules & 4052 & 46.67 & 48.70 & 65 & 2 \\
PC3\textsuperscript{$\ast$} & Molecules & 2751 & 26.36 & 28.49 & 37 & 2 \\
MOLT-4H\textsuperscript{$\ast$} & Molecules & 3977 & 46.70 & 48.74 & 65 & 2 \\
\bottomrule
\end{tabular}
\end{table}

\mypara{Graph Embedding Models}
As discussed in \autoref{sec:gnn_intro}, the graph embedding models typically consist of node embedding modules and graph pooling modules (see Section~\ref{sec:background}).
In our experiments, we use a 3-layer \sage~\cite{HYL17} module to implement node embedding.
For graph pooling, we consider the following three methods.

\begin{itemize}[leftmargin=*]
    \setlength\itemsep{-0.25em}
    \item \textbf{MeanPool~\cite{H20}.}
    Given all the node embeddings $\embed_u, \forall u \in \graph$, \meanpool directly averages all the node embeddings to obtain the graph embedding, i.e., $\embed_\graph = \frac{1}{|\graph|}\sum_{u \in \graph}\embed_u$, where $|\graph|$ is the number of nodes in \graph.
    \item \textbf{DiffPool~\cite{YYMRHL18}.}
    This is a hierarchical pooling method, which relies on multiple layers of graph pooling operations to obtain the graph embedding $\embed_\graph$.
    Concretely, we use three layers of graph pooling operations in our implementation.
    The first and second graph pooling layers narrow down the number of nodes to $0.25 \cdot |\graph|$ and $0.25^2 \cdot |\graph|$, respectively, using \diffpool operation.
    In the last layer of graph pooling, we use the mean pooling operation to generate the final graph embedding $\embed_\graph$.
    \item \textbf{MinCutPool~\cite{BGA20}.}
    This is also a hierarchical graph pooling method.
    Similar to \diffpool, we use three layers of graph pooling operations.
    The first two graph pooling layers narrow down the number of nodes to $0.5 \cdot |\graph|$ and $0.5^2 \cdot |\graph|$, respectively, using \mincutpool operation, and the last layer uses the mean pooling operation.
\end{itemize}

For presentation purpose, we use the name of graph pooling methods, namely \meanpool, \diffpool, and \mincutpool, to represent the graph embedding models in this section.

\begin{figure*}[!t]
\begin{center}
\includegraphics[width=0.8\textwidth]{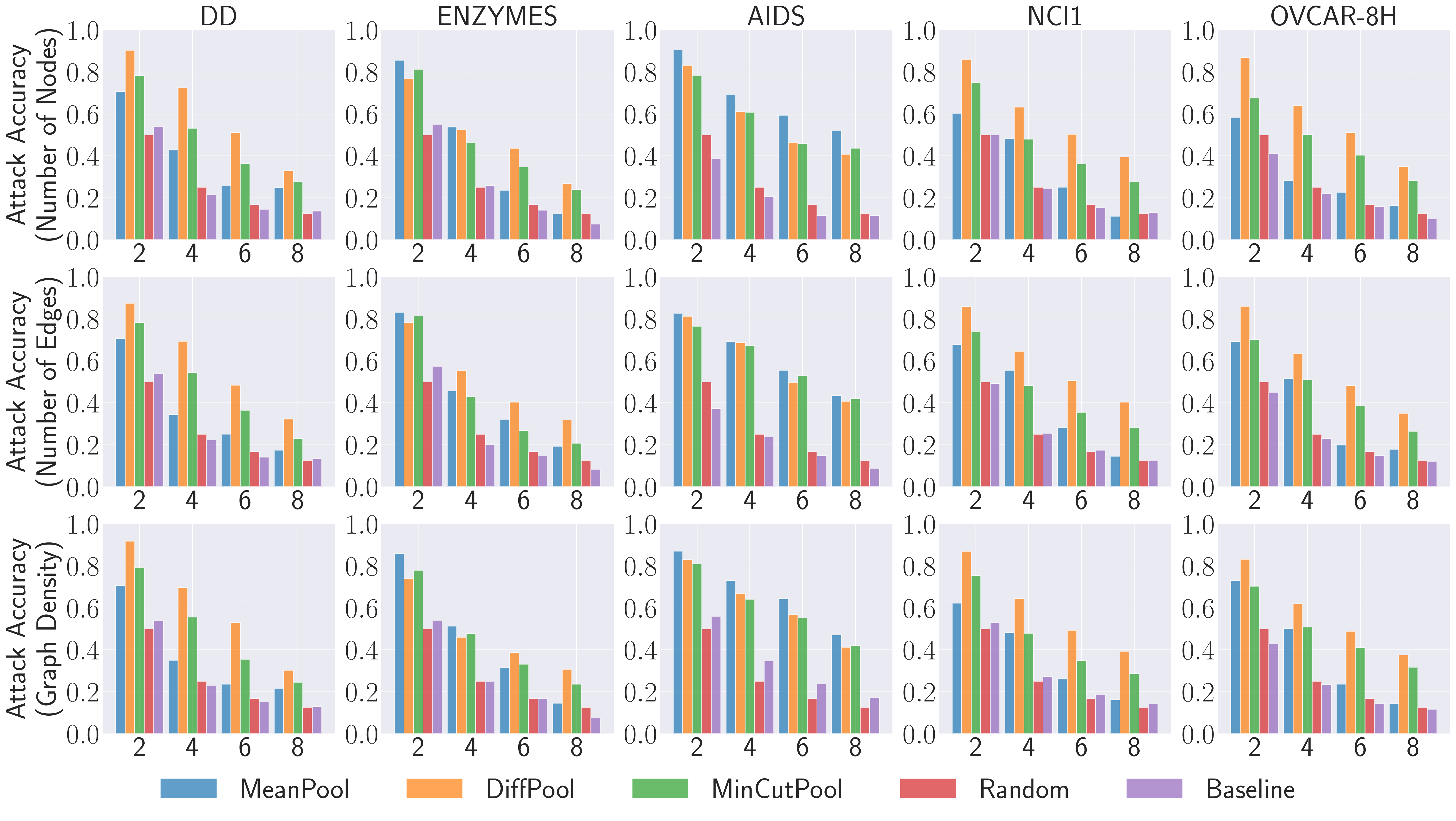}
\end{center}
\vspace{-0.4cm}
\caption{
[Higher means better attack performance.]
Attack accuracy for property inference.
Different columns represent different datasets, and different rows represent different graph properties to be inferred.
In each figure, different legends stand for different graph embedding models, different groups stand for different bucketization schemes.
The $\mathsf{Random}$ and $\mathsf{Baseline}$ method represent the random guessing and summarizing auxiliary dataset baseline, respectively.
}
\label{figure:property_inference_attack}
\end{figure*}

\mypara{Implementation}
We use the PyTorch Geometric\footnote{\url{https://github.com/rusty1s/pytorch_geometric}} library to implement all the graph embedding models.
All the attacks are implemented with Python 3.7, and conducted on an NVIDIA DGX-A100 server with 2TB memory.

\mypara{Experimental Settings}
For each dataset $\dset$, we split it into three disjoint parts, target dataset $\dset_T$, attack training dataset $\dset_A^{train}$, and attack testing dataset $\dset_A^{test}$.
The target dataset $\dset_T$ (40\%) is used to train the \targetmodelname \targetmodel, which is shared by all the three inference attacks.
The attack training dataset $\dset_A^{train}$ (30\%) corresponds to the auxiliary dataset \daux, which is used to generate the training data for the attack model.
The attack testing dataset $\dset_A^{test}$ (30\%) corresponds to the target graph \targetgraph in the attack phase.
By default, we set the graph embedding dimension $d_\embed$ as 192, which is the default setting of PyTorch Geometric.

\subsection{Property Inference Attack}
\label{subsec:evaluation_properinfer}

\begin{figure}[!tpb]
    \centering
    \begin{subfigure}{0.8\columnwidth}
    \includegraphics[width=\textwidth]{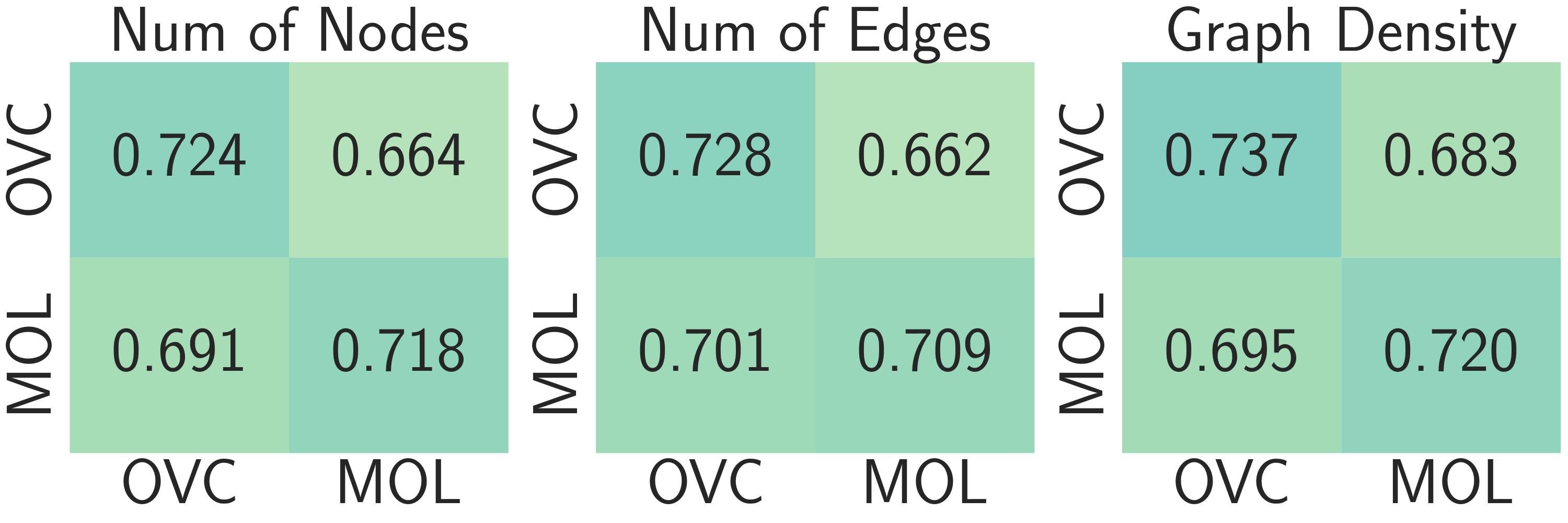}
    \includegraphics[width=\textwidth]{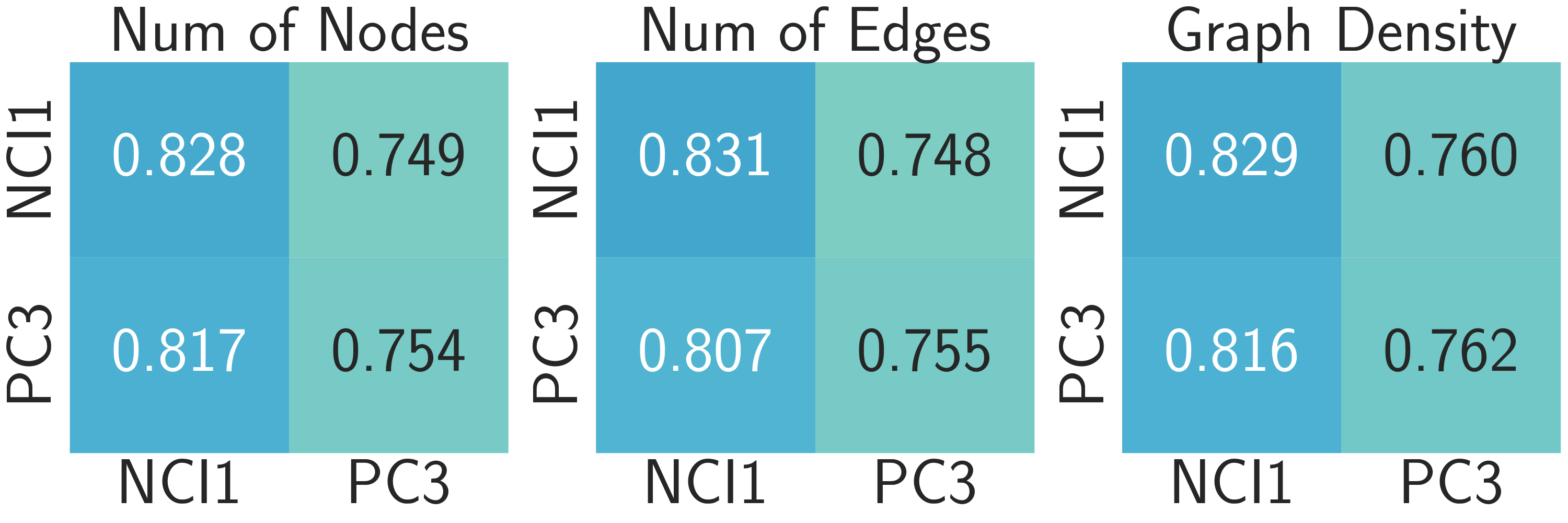}
    \label{subfigure:property_acc_dataset_transfer_2}
    \end{subfigure}
    \vspace{-0.4cm}
    \caption{
    Datasets transferability for property inference attack between OVCAR-8H (OVC) and MOLT-4H (MOL), as well as between NCI1 and PC3.
    }
    \label{figure:property_acc_dataset_transfer}
\end{figure}

\mypara{Evaluation Metrics}
As the attack goal of property inference attack is to infer the basic graph properties of the target graph \targetgraph, a commonly used metric to measure the attack performance is the \textit{attack accuracy}, which calculates the proportion of graphs being correctly inferred.

\mypara{Attack Setup}
We conduct extensive experiments on five real-world graph datasets and three state-of-the-art GNN-based graph embedding models.
In our experiments, we consider five different graph properties: Number of nodes, number of edges, graph density, graph diameter, and graph radius.
For each graph property, we bucketize its domain into $k$ bins, which transforms the attack into a multi-class classification problem.
Concretely, for the number of nodes (edges) and the graph diameter (radius), the property domain is from 1 to the maximum number of nodes (edges) and the maximum graph diameter (radius) in the auxiliary dataset \daux.
For the graph density, the property domain is $[0.0, 1.0]$.
In our experiments, we consider four different \textit{bucketization schemes}, i.e., $k \in \{2,4,6,8\}$.

\mypara{Competitors} 
To validate the effectiveness of our proposed attack, we need to compare with two baseline attacks.
\begin{itemize}[leftmargin=*]
    \setlength\itemsep{-0.2em}
    \item \textbf{Random Guessing (Random).}
    The most straightforward baseline is random guessing,
    which varies for different bucketization schemes.
    For instance, the attack accuracy of random guessing for $k=2$ and $k=8$ are $0.5$ and $0.125$.
    \item \textbf{Directly Summarizing the Auxiliary Dataset (Baseline).}
    Another baseline attack is directly summarizing the properties from the auxiliary dataset \daux instead of training a classifier.
    Concretely, we calculate the average property values from \daux, and use them for predicting the properties of the target graphs.
\end{itemize}

\begin{figure*}[!t]
\centering
\begin{subfigure}{1.8\columnwidth}
\includegraphics[width=1\columnwidth]{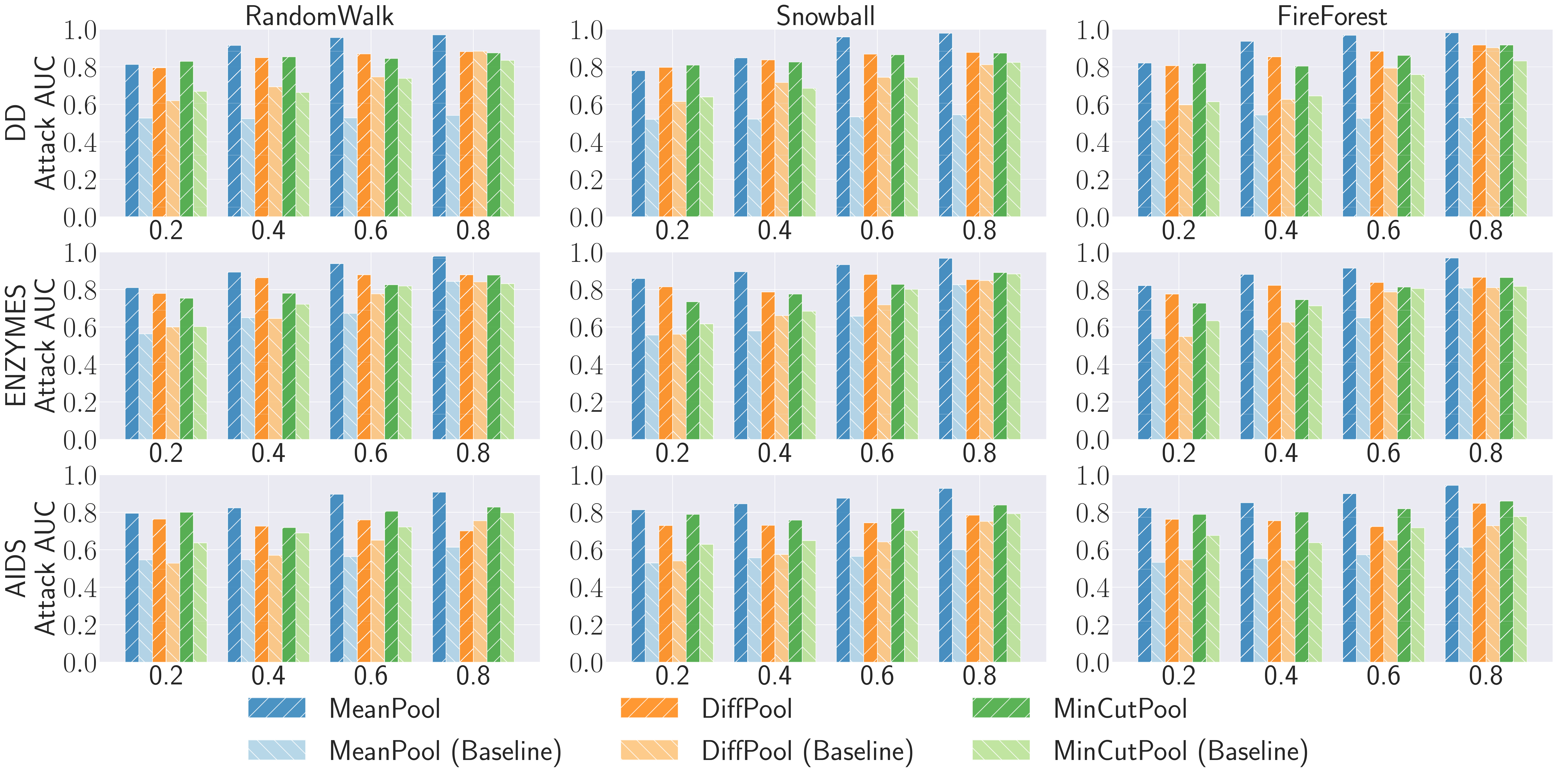}
\end{subfigure} \\
\caption{
[Higher means better attack performance.] 
Attack AUC for subgraph inference attack.
Different rows and columns represent different datasets and graph sampling methods.
In each figure, different legends and groups stand for different graph embedding models and different sampling ratios.
We use element-wise difference method to generate the feature vector \subfeat.
}
\label{figure:subgraph_inference_attack_compare}
\end{figure*}

\mypara{Experimental Results}
\autoref{figure:property_inference_attack} illustrates the attack performance, where different rows represent different graph properties, and different columns represent different datasets.
Due to space limitation, we defer the results of graph diameter and graph radius to \autoref{appsub:additional_property_infer}.

In general, the experimental results show that our attack outperforms two baseline attacks in most of the settings.
For instance, when the bucketization scheme $k=2$, on the number of nodes property, we can achieve an attack accuracy of $0.904$ on the DD dataset for the \diffpool model, while the attack accuracy of random guessing and summarizing auxiliary dataset baseline is $0.500$ and $0.541$, respectively.
We further observe that a larger bucketization scheme $k$ leads to worse attack accuracy.
This is expected because larger $k$ requires higher granularity of graph structural information, and is more difficult for the classifier to distinguish.
In addition, we note that, in most of the cases, the attack accuracy on the \meanpool model is worse than that of the other two graph embedding models, and sometimes even close to that of the random guessing baseline.
This can be explained by the fact that the \meanpool model directly averages all the node embeddings, which might lose some graph structural information.

\mypara{Datasets Transferability}
In previous experiments, we assume the auxiliary dataset \daux comes from the same distribution as the target graphs.
To relax this assumption, we conduct additional experiments when \daux comes from different distribution than the target graphs.
We evaluate the transferability between OVCAR-8H (OVC) and MOLT-4H (MOL), as well as between NCI1 and PC3 on \mincutpool with $k=2$.
The experimental results in \autoref{figure:property_acc_dataset_transfer} show that our property inference attack is still effective when \daux and the target graphs come from different distributions.

\begin{table*}[!t]
\centering
\caption{
Attack AUC for different feature construction methods in subgraph inference attack.
The graph embedding model is \diffpool and the graph sampling method is \rws.
Due to space limitation, we use Concat, EDist, and EDiff to represent Concatenation, Euclidean Distance, and Element-wise Difference, respectively.
}
\label{table:subgraph_infer_feature_construction}
\footnotesize
\setlength{\tabcolsep}{0.3em}
\renewcommand{\arraystretch}{1.0}
\scriptsize
\begin{tabular}{c | c c c | c c c | c c c | c c c }
\toprule
& \multicolumn{3}{c|}{0.8} & \multicolumn{3}{c|}{0.6} & \multicolumn{3}{c|}{0.4} & \multicolumn{3}{c}{0.2}\\
Dataset & Concat & EDist & EDiff & Concat & EDist & EDiff & Concat & EDist & EDiff & Concat & EDist & EDiff \\
\toprule
DD & 0.53 $\pm$ 0.01 & 0.81 $\pm$ 0.06 & \textbf{0.88 $\pm$ 0.01} & 0.51 $\pm$ 0.01 & 0.79 $\pm$ 0.04 & \textbf{0.87 $\pm$ 0.01} & 0.52 $\pm$ 0.01 & 0.79 $\pm$ 0.02 & \textbf{0.85 $\pm$ 0.01} & 0.50 $\pm$ 0.02 & 0.71 $\pm$ 0.08 & \textbf{0.80 $\pm$ 0.00} \\
\midrule
ENZYMES & 0.49 $\pm$ 0.02 & 0.63 $\pm$ 0.10 & \textbf{0.88 $\pm$ 0.03} & 0.52 $\pm$ 0.03 & 0.71 $\pm$ 0.10 & \textbf{0.88 $\pm$ 0.03} & 0.54 $\pm$ 0.02 & 0.56 $\pm$ 0.07 & \textbf{0.86 $\pm$ 0.01} & 0.48 $\pm$ 0.02 & 0.53 $\pm$ 0.03 & \textbf{0.78 $\pm$ 0.01} \\
\midrule
AIDS & 0.51 $\pm$ 0.01 & 0.53 $\pm$ 0.04 & \textbf{0.78 $\pm$ 0.04} & 0.55 $\pm$ 0.01 & 0.51 $\pm$ 0.02 & \textbf{0.76 $\pm$ 0.05} & 0.54 $\pm$ 0.01 & 0.51 $\pm$ 0.03 & \textbf{0.73 $\pm$ 0.06} & 0.56 $\pm$ 0.02 & 0.50 $\pm$ 0.00 & \textbf{0.76 $\pm$ 0.05} \\
\midrule
NCI1 & 0.51 $\pm$ 0.00 & 0.51 $\pm$ 0.02 & \textbf{0.70 $\pm$ 0.06} & 0.49 $\pm$ 0.02 & 0.52 $\pm$ 0.01 & \textbf{0.67 $\pm$ 0.06} & 0.50 $\pm$ 0.01 & 0.51 $\pm$ 0.01 & \textbf{0.64 $\pm$ 0.03} & 0.49 $\pm$ 0.01 & 0.51 $\pm$ 0.01 & \textbf{0.64 $\pm$ 0.00} \\
\midrule
OVCAR-8H & 0.54 $\pm$ 0.01 & 0.63 $\pm$ 0.12 & \textbf{0.89 $\pm$ 0.02} & 0.50 $\pm$ 0.04 & 0.69 $\pm$ 0.09 & \textbf{0.88 $\pm$ 0.02} & 0.51 $\pm$ 0.03 & 0.74 $\pm$ 0.02 & \textbf{0.84 $\pm$ 0.01} & 0.54 $\pm$ 0.01 & 0.60 $\pm$ 0.13 & \textbf{0.82 $\pm$ 0.02} \\
\bottomrule
\end{tabular}
\end{table*}

\subsection{Subgraph Inference Attack}
\label{subsec:evaluation_subinfer}

\mypara{Evaluation Metrics}
Recall that the subgraph inference attack is a binary classification task; thus we use the \textit{AUC} metric to measure the attack performance, which is widely used to measure the performance of binary classification in a range of thresholds~\cite{FLJLPR14,BHPZ17,PTC18,JSBZG19,ZHSMVB20,CZWBHZ21}.
The higher AUC value implies better attack performance. 
An AUC value of 1 implies maximum performance (true-positive rate of 1 with a  false-positive rate of 0) while an AUC value of 0.5 means performance equivalent to random guessing.

\mypara{Attack Setup}
We conduct extensive experiments on five graph datasets and three graph embedding models to evaluate the effectiveness of our proposed attack.
To obtain the subgraph, we rely on three graph sampling methods: 

Random walk sampling, snowball sampling, and forest fire sampling.
We refer the readers to \autoref{app:experimental_details} for detailed descriptions of these sampling methods.
For each sampling method, we consider four different \textit{sampling ratios}, i.e., $\{0.2,0.4,0.6,0.8\}$, which determines how many nodes are contained in the subgraph.
In practice, the sampling ratio is determined by the size of the subgraph of interest.

We use element-wise difference method to generate the feature vector \subfeat.
We generate the same number of positive samples and negative samples in both training and testing datasets to learn balanced model.

\mypara{Competitor}
Recall that we integrate a graph embedding extractor in the attack model to transform the subgraph into subgraph embedding in \autoref{sec:subgraph_infer}.
The embedding extractor is jointly trained with the binary classifier in the attack model.
An alternative for subgraph inference is to generate the subgraph embedding from the target model together with the target graph embedding, and then train an isolated binary classifier as attack model.
To validate the necessity of integrating embedding extractor in the attack model, we compare with the baseline attack that obtains subgraph embeddings from the target model.

\begin{figure}[!tpb]
    \centering
    \begin{subfigure}{0.75\columnwidth}
    \includegraphics[width=\textwidth]{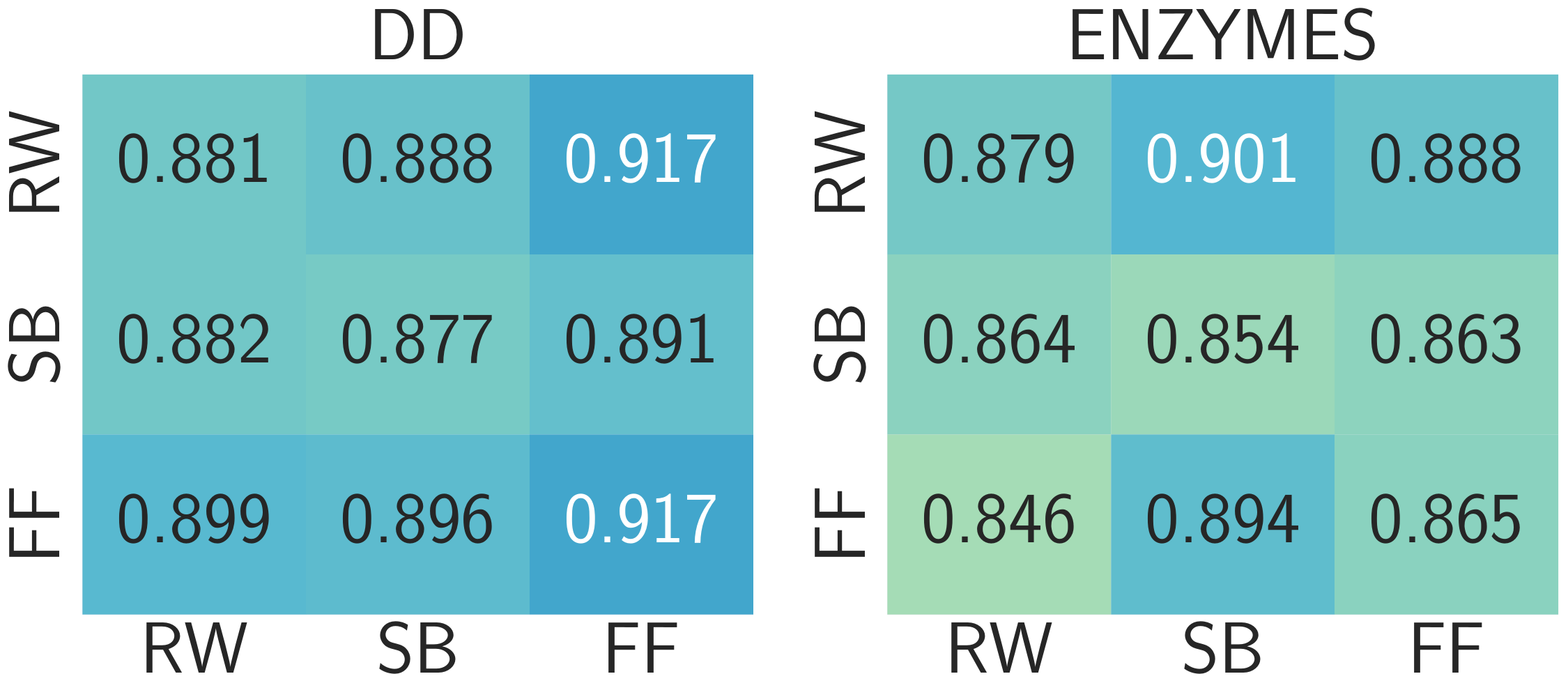}
    \label{subfigure:sampling_method_transfer0.8}
    \end{subfigure}
    \vspace{-0.4cm}
    \caption{
    Sampling methods transferability results for subgraph inference attack.
    RW, SB, and FF are abbreviations for \rws, \sbs, and \ffs, respectively.
    }
    \label{figure:sampling_method_transfer}
\end{figure}

\mypara{Experimental Results}
\autoref{figure:subgraph_inference_attack_compare} illustrates the attack performance, where different rows represent different datasets, and different columns represent different sampling methods.
Due to space limitation, we defer the results of other datasets to \autoref{appsub:additional_subgraph_infer}.
The experimental results show that our attack is effective in most of the settings, especially when the sampling ratio is $0.8$.
For instance, we can achieve $0.982$ attack AUC on the DD dataset and \meanpool model with \ffs sampling method.
Besides, we observe that when the sampling ratio decreases, the attack AUC decreases for most of the settings.
This is expected as the positive samples and the negative samples tend to be more similar to each other on smaller subgraphs, making the attack model more difficult to distinguish between them.
Despite this, our attack can still achieve $0.859$ attack AUC on ENSYMES and \meanpool with \sbs when the sampling ratio is 0.2.

Comparing different graph embedding models, we further observe that the subgraph inference attack performs the best on the \meanpool model in most of the settings, which is opposite to the property inference attack.
We suspect this is because \diffpool and \mincutpool decompose the graph structure during their pooling process; thus, the subgraph as a whole might never be seen by the target model.  
This makes it harder for graph embedding matching to be effective.

\mypara{Necessity of Embedding Extractor}
Comparing with the baseline, we observe that our subgraph inference attack consistently outperforms the baseline attack in most of the cases, especially when the sampling ratio is small.
For instance, on the DD dataset, when the sampling ratio is 0.2, our attack achieves 0.821 AUC on \meanpool model and \ffs sampling method, while the baseline attack achieves AUC of 0.515.
We further observe that when the sampling ratio increases, the baseline attack can gradually achieve comparable attack AUC as our attack.
This is expected as distinguishing between the positive subgraph and negative subgraph is much easier when the sampling ratio is large.

\begin{figure}[!tpb]
    \centering
    \begin{subfigure}{0.75\columnwidth}
    \includegraphics[width=\textwidth]{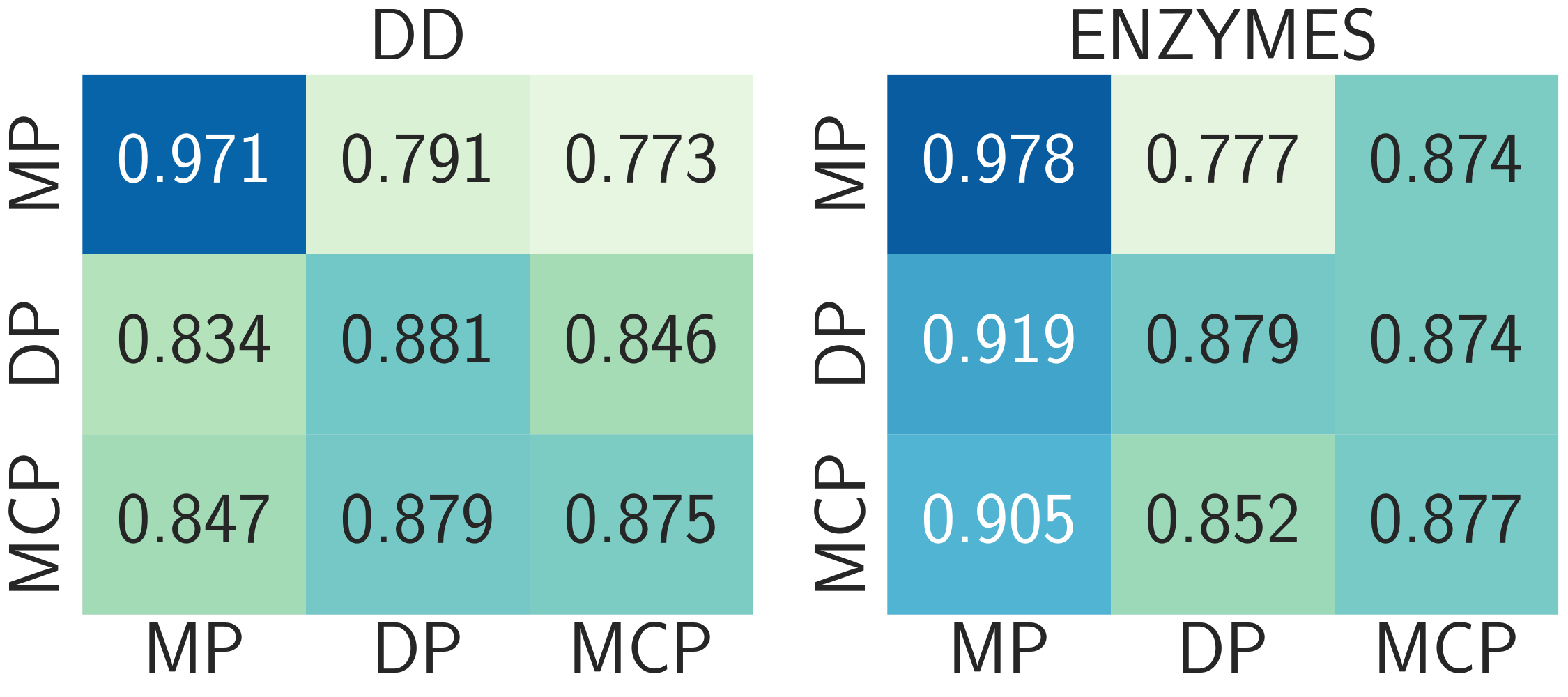}
    \label{subfigure:model_transfer_0.8}
    \end{subfigure}
    \vspace{-0.4cm}
    \caption{
    Embedding models transferability for subgraph inference attack.
    MP, DP, and MCP are abbreviations for \meanpool, \diffpool, and \mincutpool, respectively.
    }
    \label{figure:model_transfer}
\end{figure}

\mypara{Comparison of Feature Construction Methods}
We propose three strategies to aggregate the graph embeddings of the target graph and the subgraph of interest in the attack model \subinfermodel, namely concatenation, element-wise difference, and Euclidean distance, in \autoref{sec:subgraph_infer}.
We now compare the performance of different strategies.
\autoref{table:subgraph_infer_feature_construction} shows the experimental results on five datasets when the graph embedding model is \diffpool and the graph sampling method is \rws.

We observe that the element-wise difference method achieves the best performance, while the concatenation method has an attack AUC close to random guessing.
This indicates that the discrepancy information between two graph embeddings (element-wise difference method) is more informative than the plain graph embeddings (concatenation method) in terms of subgraph inference attack.
Note that the Euclidean distance also implicitly captures the discrepancy information of two graph embeddings, while it relies on one scalar value and loses other rich discrepancy information.

\mypara{Sampling Methods Transferability}
So far, our experiments use the same sampling method for the auxiliary graph to train the attack model and the target graph to test the attack model.
We conduct additional experiments to show whether our attack still works when the sampling methods are different.
\autoref{figure:sampling_method_transfer} illustrates the experimental results on DD and ENZYMES datasets.
We use \diffpool as the graph embedding model and adopt a sampling ratio of $0.8$.
As we can see, in most cases, the sampling methods do not have a significant impact on the attack performance.

\mypara{Embedding Models Transferability}
In previous experiments, the architecture of the graph embedding extractor in the attack model is the same as the \targetmodelname.
In practice, the model architecture of the \targetmodelname might be unknown to the adversaries.
To understand whether our attack still works when the architectures are different, we conduct experiments on the DD and ENZYMES datasets.
\autoref{figure:model_transfer} illustrates the experimental results of \rws sampling method with a sampling ratio of $0.8$.
We observe that the attack performance slightly drops when the model architectures are different.
Despite this, we can still achieve 0.773 attack AUC in the worse case.

\begin{figure}[!tpb]
    \centering
    \begin{subfigure}{0.8\columnwidth}
    \includegraphics[width=\textwidth]{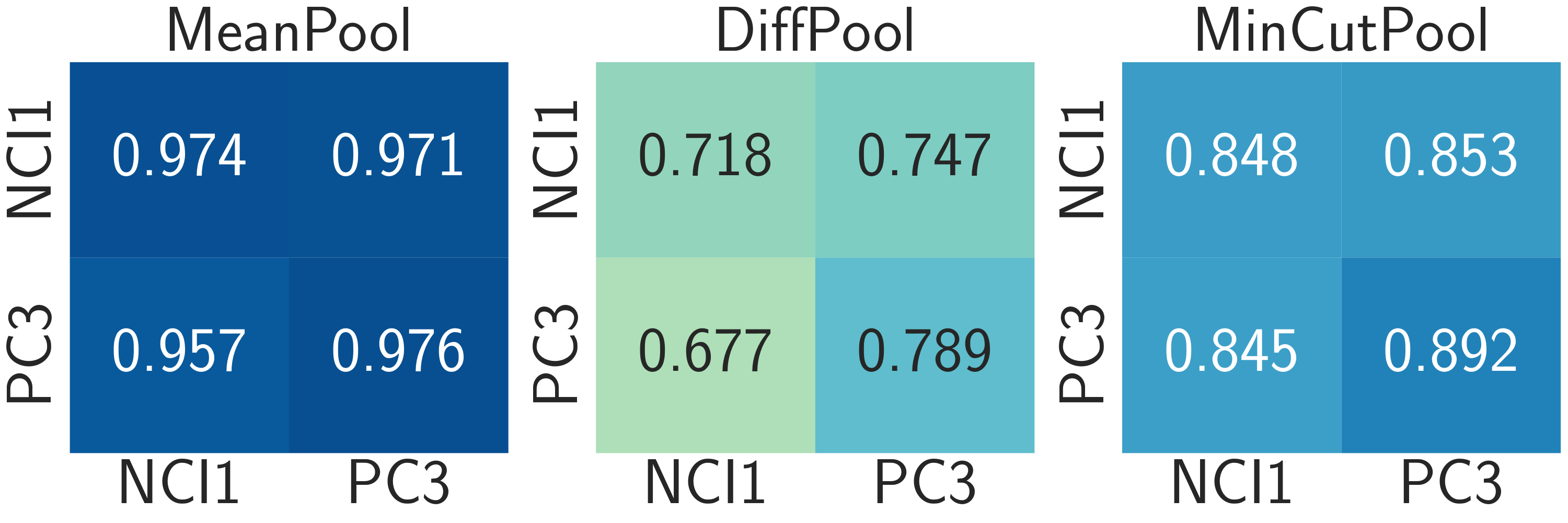}
    \includegraphics[width=\textwidth]{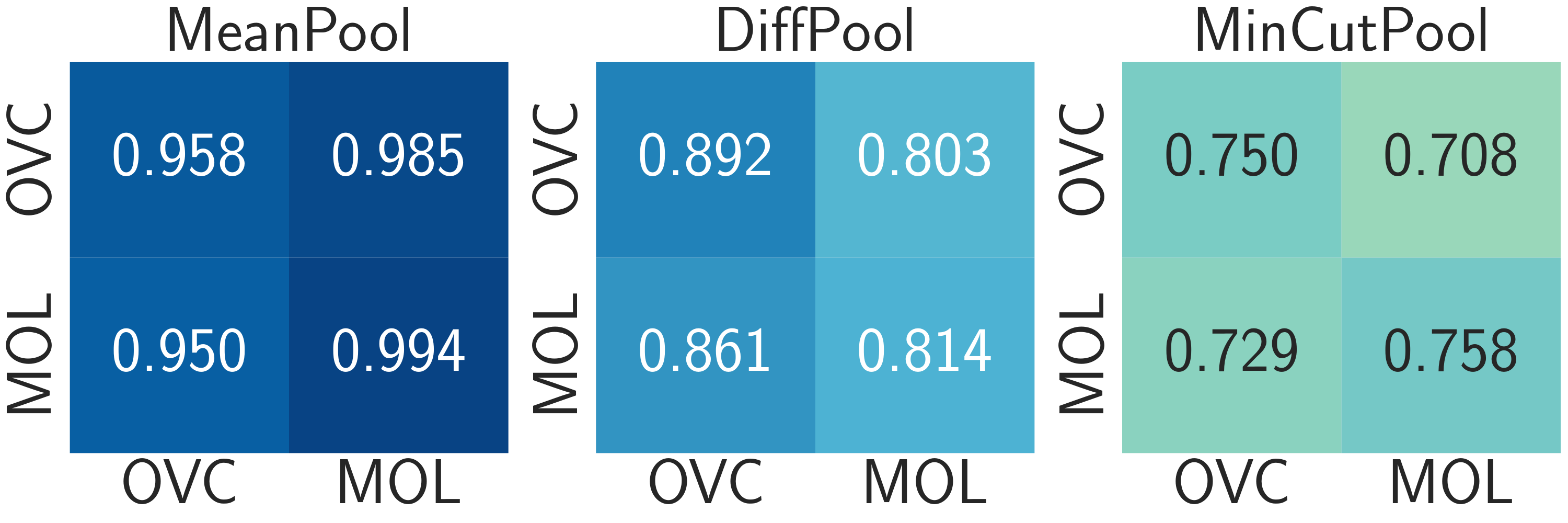}
    \label{subfigure:dataset_transfer0.8}
    \end{subfigure} %
    \vspace{-0.4cm}
    \caption{
    Dataset transferability for subgraph inference between OVCAR-8H (OVC) and MOLT-4H (MOL), as well as between NCI1 and PC3.}
    \label{figure:subgraph_dataset_transfer}
\end{figure}

\mypara{Datasets Transferability}
Similar to property inference attack, to relax the assumption that \daux comes from the same distribution of the target graphs, we conduct additional experiments when \daux and the target graphs come from different distributions.
We experiment on the \rws method with a sampling ratio of 0.8.
The experimental results in \autoref{figure:subgraph_dataset_transfer} show that our subgraph inference attack is still effective for dataset transfer.

\subsection{Graph Reconstruction Attack}
\label{subsec:evaluation_graph_recon}

\mypara{Evaluation Metrics}
We evaluate the performance of graph reconstruction from two perspectives: 

\begin{itemize}[leftmargin=*]
    \setlength\itemsep{-0.2em}
    \item \textbf{Graph Isomorphism}.
    The graph isomorphism compares the structure of the reconstructed graph \recongraph with the target graph \targetgraph, and determines their similarity.
    The graph isomorphism problem is well-known to be intractable in polynomial time; thus, approximate algorithms such as \textit{Weisfeiler-Lehman (WL) algorithm} are widely used for addressing it~\cite{SSLMB11,XHLJ19,MRFHLRG19}.
    The general idea of WL algorithm is to iteratively calculate the WL graph kernel of two graphs.
    We normalize the WL graph kernel in the range of [0.0, 1.0], and a WL graph kernel of 1.0 means two graphs perfectly match.
    We adopt the DGL implementation of WL algorithm in our experiments.\footnote{\url{https://github.com/InkToYou/WL-Kernel-DGL}}

    \item \textbf{Macro-level Graph Statistics}.
    Recall that the objective of the graph reconstruction attack is to generate a graph \recongraph that has similar graph statistics with the target graph \targetgraph.
    In practice, there are a plethora of graph structural statistics to analyze a graph.
    In this paper, we adopt four widely used graph statistics: 
    Degree distribution, local clustering coefficient (LCC), betweenness centrality (BC), and closeness centrality (CC).
    We refer the readers to \autoref{app:experimental_details} for detailed descriptions of these statistics.
\end{itemize}

Note that the number of nodes in \recongraph might be different from the target graph \targetgraph due to the graph auto-encoder architecture, and there are no node orderings imposed for \recongraph and \targetgraph; thus we cannot directly compare the node-level graph statistics including LCC, CC, and BC.
To address this issue, we bucketize the statistic domain into 10 bins and measure their distributions.
For each graph statistic, we use three metrics to measure the distribution similarity between the target graph \targetgraph and the reconstructed graph \recongraph: \textit{Cosine similarity}, \textit{Wasserstein distance}, and \textit{Jensen-Shannon (JS) divergence}.
Intuitively, higher cosine similarity and lower Wasserstein distance/JS divergence mean better attack performance.
The ranges of cosine similarity, Wasserstein distance, and JS divergence are $[-1.0, 1.0]$, $[0.0, 1.0]$, and $[0.0, 1.0]$, respectively.

\mypara{Attack Setup}
Recall that both space and time complexity of the graph matching algorithm are $O(n^4)$,
we conduct our experiments on three small datasets in \autoref{table:dataset_statistic}, i.e., AIDS, ENZYMES, and NCI1, and three graph embedding models.
We run all the experiments five times with the mean and standard deviation reported.

\begin{table}
\centering
\caption{[Higher means better attack performance.]
Attack performance of graph reconstruction measured by graph isomorphism.
}
\label{table:graph_recon_isomophism}
\footnotesize
\setlength{\tabcolsep}{0.6em}
\renewcommand{\arraystretch}{1.0}
\begin{tabular}{c | c  c  c }
\toprule
Dataset
& \multicolumn{1}{c}{\diffpool} & \multicolumn{1}{c}{ \meanpool} & \multicolumn{1}{c}{\mincutpool}\\
\toprule
\multirow{1}{*}{AIDS}
& 0.875 $\pm$ 0.003  & 0.794 $\pm$ 0.003 & 0.869 $\pm$ 0.002 \\
\multirow{1}{*}{ENZYMES}& 0.670 $\pm$ 0.019 & 0.653 $\pm$ 0.022 & 0.704 $\pm$ 0.012 \\
\multirow{1}{*}{NCI1}
& 0.752 $\pm$ 0.005 & 0.771 $\pm$ 0.010 & 0.693 $\pm$ 0.007 \\
\bottomrule
\end{tabular}
\end{table}

\mypara{Experimental Results}
\autoref{table:graph_recon_isomophism} and \autoref{table:graph_recon} illustrate the attack performance in terms of graph isomorphism and macro-level graph statistics (measured by cosine similarity), respectively.
Due to space limitation, we defer the results of the macro-level graph statistics measured by Wasserstein distance and JS divergence to \autoref{appsub:additional_graph_recon}.
In general, our attack achieves strong performance.
For instance, the WL graph kernel on AIDS and \diffpool achieves 0.875.
Besides, the cosine similarity of the betweenness centrality distribution is larger than 0.85 for all the settings.
We can also achieve 0.99 cosine similarity for local clustering coefficient distribution for the AIDS and NCI1 datasets.
For degree distribution and closeness centrality distribution, the attack performance is slightly worse; however, we can still achieve cosine similarity larger than or close to 0.5.

\smallskip
To investigate the impact of the quality of the auto-encoder on the attack performance, we conduct additional experiments on the auto-encoders trained in  different epochs.
Due to space limitation, we defer the experimental results to \autoref{appsub:additional_graph_recon}.

\section{Defenses}
\label{sec:defense}

\mypara{Graph Embedding Perturbation}
A commonly used defense mechanism for inference attacks is adding perturbation to the output of the model~\cite{ZWHLBHCZ21}.
In this paper, we propose to add perturbations to the target graph embedding \targetembed to defend our proposed inference attacks.
Formally, given the target graph embedding \targetembed, the data owner only shares a noisy version of graph embedding $\Tilde{\targetembed} = \targetembed + \Lapp{\beta}$ to the third party, where $\Lapp{\beta}$ denotes a random variable sampled from the Laplace distribution with scale parameter $\beta$; that is, $\Pr{\Lapp{\beta}=x} = \frac{1}{2\beta} e^{-|x|/\beta}$.
Notice that adding noise to the graph embedding vector may destroy the graph structural information, thus affect the normal tasks such as graph classification.
Therefore, we need to choose a moderate level of noise to tradeoff the defense effectiveness and the performance of the normal tasks.

\begin{table}
\centering
\caption{[Higher means better attack performance.]
Attack performance of graph reconstruction measured by macro-level graph statistics, the similarity of which is measured by cosine similarity.
}
\label{table:graph_recon}
\footnotesize
\setlength{\tabcolsep}{0.2em}
\renewcommand{\arraystretch}{1.4}
\scriptsize
\begin{tabular}{c c | c  c  c  c}
\toprule
Dataset & Target Model & \multicolumn{1}{c}{Degree Dist.} & \multicolumn{1}{c}{LCC Dist.} & \multicolumn{1}{c}{BC Dist.} & \multicolumn{1}{c}{CC Dist.}\\
\toprule
\multirow{3}{*}{\rotatebox[origin=c]{90}{AIDS}}
& \meanpool & 0.651 $\pm$ 0.001 & 0.999 $\pm$ 0.001 & 0.987 $\pm$ 0.001 & 0.876 $\pm$ 0.002 \\
& \diffpool & 0.894 $\pm$ 0.001 & 0.999 $\pm$ 0.001 & 0.983 $\pm$ 0.001 & 0.787 $\pm$ 0.002 \\
& \mincutpool & 0.888 $\pm$ 0.003 & 0.999 $\pm$ 0.001 & 0.983 $\pm$ 0.001 & 0.785 $\pm$ 0.006 \\
\midrule
\multirow{3}{*}{\rotatebox[origin=c]{90}{ENZYMES}}
& \meanpool & 0.450 $\pm$ 0.070 & 0.646 $\pm$ 0.005 & 0.959 $\pm$ 0.001 & 0.516 $\pm$ 0.037 \\
& \diffpool & 0.519 $\pm$ 0.007 & 0.661 $\pm$ 0.008 & 0.958 $\pm$ 0.001 & 0.504 $\pm$ 0.005 \\
& \mincutpool & 0.467 $\pm$ 0.019 & 0.490 $\pm$ 0.009 & 0.916 $\pm$ 0.001 & 0.414 $\pm$ 0.009 \\
\midrule
\multirow{3}{*}{\rotatebox[origin=c]{90}{NCI1}}
& \meanpool & 0.736 $\pm$ 0.003 & 0.999 $\pm$ 0.001 & 0.877 $\pm$ 0.001 & 0.402 $\pm$ 0.001 \\
& \diffpool & 0.633 $\pm$ 0.002 & 0.999 $\pm$ 0.001 & 0.877 $\pm$ 0.001 & 0.495 $\pm$ 0.002 \\
& \mincutpool & 0.570 $\pm$ 0.002 & 0.999 $\pm$ 0.001 & 0.877 $\pm$ 0.001 & 0.496 $\pm$ 0.001 \\
\bottomrule
\end{tabular}
\end{table}

\begin{figure*}[!htpb]
\begin{center}
\includegraphics[width=0.9\textwidth]{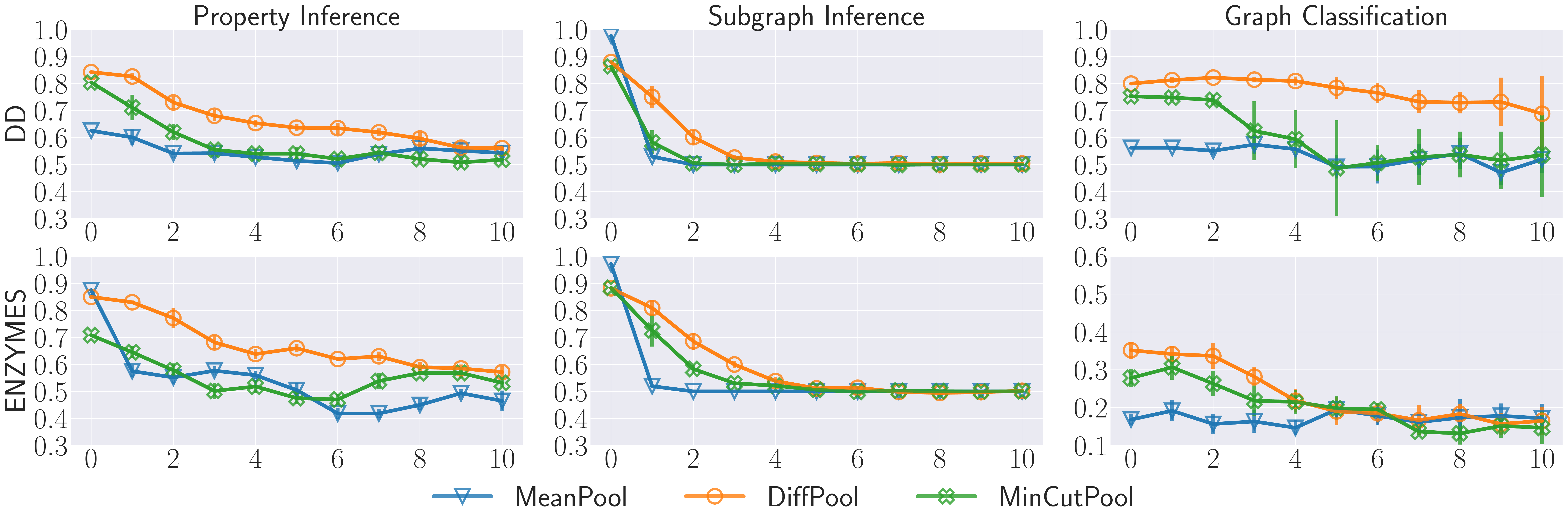}
\end{center}
\vspace{-0.4cm}
\caption{
Graph embedding perturbation defense on the DD and ENZYMES datasets (different rows).
The first two columns represent the attack performance of property inference and subgraph inference respectively, the last column represents the accuracy of normal graph classification task.
In each figure, the x-axis stands for the scaling parameter $\beta$ for Laplace noise, where larger $\beta$ means higher noise level.
The y-axis stands for the attack performance/normal graph classification accuracy.
}
\label{figure:defense}
\end{figure*}

\mypara{Defense Evaluation Setup}
We conduct experiments to validate the effectiveness of our proposed defense against all the inference attacks, as well as the impact on normal graph classification task.
For property inference attack, we evaluate the performance of graph density with bucketization scheme $k=2$.
For subgraph inference attack, we consider the \rws sampling method with sampling ratio of $0.8$.
We conduct our experiments on DD and ENZYMES datasets and three graph embedding models.
Due to space limitation, we refer the readers to \autoref{appsub:additional_defense} for experimental results for other datasets and graph reconstruction attack.

\mypara{Defense Evaluation Results}
\autoref{figure:defense} illustrates the experimental results, where the first and second column represents the attack performance of property inference attack and subgraph inference attack respectively, the last column represents the accuracy of the normal graph classification task.
In each figure, the x-aixs stands for the scaling parameter $\beta$ of Laplace noise, where larger $\beta$ means larger noise.
We observe that when the noise level increases, the attack performance for both property inference and subgraph inference attack decreases.
This is expected since more noise will hide more structural information contained in the graph embedding.
On the other hand, the accuracy of the graph classification tasks will also decrease when the noise level increase.
To defend against the inference attacks while preserving the utility for normal tasks, one needs to carefully choose the noise level.
For instance, when we set the standard deviation of Laplace noise to 2, the  performance of subgraph inference attack significantly drops while the graph classification accuracy only slightly decreases.

\section{Related Work}
\label{sec:related}
In this section, we review the research work close to our proposed attacks.
We refer the readers to~\cite{GF18,ZCZ20} for in-depth overview of different GNN models, and~\cite{DLTHWZS18,SDYWYHL18,JLXWT20,XMLDLTJ20} for comprehensive surveys of existing adversarial attacks and defense strategies on GNNs.

\mypara{Causative Attacks on GNNs} 
Causative attack allows attackers to manipulate training dataset in order to change the parameters of the target model. 
In the context of causative attacks on GNNs, Z{\"u}gner~\etal~\cite{ZAG18} was the first research work that introduced unnoticeable adversarial perturbations targeting the node’s features and the graph structure to reduce the accuracy of node classification via graph convolutional networks. 
Following this direction, researchers investigated different adversarial attack strategies (\ie edge/node-level/structure/attribute perturbation) to achieve various attack objectives, such as reducing the accuracy of node classification~\cite{BG192,XCLCWHL19,WWTDLZ19,EADP20,SWTHH20,MDM20}, link prediction~\cite{BG192,LJL20}, graph classification~\cite{DLTHWZS18,XPJW21}, etc.
Our attacks do not tamper with the training data that is used to construct the GNN models.

\mypara{Exploratory Attacks on GNNs}
Exploratory attack does not change the parameters of the target model. 
Instead, the attacker sends new data to the target model and observes the model’s decisions on these carefully crafted input data.
However, graph-based machine learning under adversarial exploratory setting is much less explored.
In particular, only a few studies~\cite{HJBGZ21,WYPY20,DBS20} focused on exploratory attacks on GNNs.
For instance, He~\etal~\cite{HJBGZ21} proposed link stealing attack to infer, from the outputs of a GNN model, whether there exists a link between any pair of nodes in the graph used to train the model. 
Wu~\etal~\cite{WYPY20} discussed GNN model extraction attack, given various levels of background knowledge, by gathering both the input-output query pairs and the graph structure to reconstruct a duplicated model.
Duddu~\etal~\cite{DBS20} proposed a graph reconstruction attack against \textit{node embeddings}; however, there are several difference from our graph reconstruction attack.
First, the task is different, \cite{DBS20} aims to reconstruct a graph from a set of node embeddings, while ours is to reconstruct the graph from a graph embedding.
Also, the node embeddings targeted by \cite{DBS20} are generated from traditional node embedding method such as Deepwalk~\cite{PAS14} and node2vec~\cite{GL16}, while ours focus on state-of-the-art GNN.
In addition, our threat model is more general and practical as we are only given one embedding vector of the target graph instead of all embeddings of all the nodes. 
In this sense, our adversary has much less background knowledge than that of \cite{DBS20}.
Besides, their method uses the non-learnable dot product as the decoder.
Our approach leverages a learnable decoder and can be further fine-tuned to enhance graph reconstruction performance.

\mypara{Defense of Adversarial Attacks on GNNs}
The emerging attacks on GNNs leads to an arm race. 
To mitigate those attacks, several defense strategies (\eg graph sanitization~\cite{WWTDLZ19}, adversarial training~\cite{DSZLW19,FHTC19} and certification of robustness~\cite{BG19}) have been proposed.
One important direction of those defense strategies is to reduce the sensitivity of GNNs via adversarial training so that the train GNNs are robust to structure perturbation~\cite{DSZLW19} and attribution perturbation~\cite{FHTC19}.
Beside, robustness certification~\cite{BG19} is an emerging research direction that measure and reason the safety of graph neural networks  under adversarial perturbation.
Note that aforementioned defense mechanisms focus on mitigating causative attacks on GNNs, hence they are are not design to protect GNNs from exploratory attacks. 

\section{Conclusion}
\label{sec:conclusion}
In this paper, we investigate the information leakage of graph embedding.
Concretely, we propose three different attacks to extract information from the target graph given the graph embedding.
First, we can successfully infer graph properties, such as the number of nodes, the number of edges, and graph density, of the target graph.
Second, given a subgraph of interest and the graph embedding, we can determine with high confidence that whether the subgraph is contained in the target graph.
Third, we propose a novel graph reconstruction attack that can reconstruct a graph that has similar graph statistics with the target graph.
We further propose an embedding perturbation based defense to mitigate the inference attacks without noticeable accuracy degradation.

\section*{Acknowledgments} 
We thank the anonymous reviewers for their constructive feedback.
This work is partially funded by the Helmholtz Association within the project ``Trustworthy Federated Data Analytics'' (TFDA) (funding number ZT-I-OO1 4).

\balance
{
\bibliographystyle{plain}
\bibliography{normal_generated_py3.bib}
}
\appendix

\section{Notations}
\label{app:notations}

The frequently used notations used in this paper is summarized in \autoref{table:notations}.

\begin{table}[!hb]
\caption{Summary of the notations used in this paper.}
\centering
\vspace{-0.1cm}
\resizebox{0.9\linewidth}{!} 
{
\begin{tabular}{l|l}
\toprule 
\textbf{Notation} & \textbf{Description} \\ \toprule
	  $\graph = \tuple{\nodeset, \adj, \feat}$   & Graph   \\ 
      $u, v \in \nodeset$ & Nodes in \graph \\ 
      $n = |\nodeset|$ & Number of nodes \\
      $d_{\feat}$ / $d_{\embed}$ & Dimension of attributes / embeddings \\ 
      $\adj \in \{ 0, 1\}^{n \times n}$ & Adjacency matrix of \graph \\
      $\feat \in \mathbb{R}^{n \times d_{\feat}}$ &  Attributes associated with \nodeset \\ 
      \neigh{u} & Neighborhood nodes of $u$ \\ 
      $\subgraph$ & Subgraph of \graph \\
	  $\targetgraph$ / $\auxgraph$ & Target / auxiliary graph \\
	  \daux & Auxiliary dataset ($\auxgraph \in \daux$) \\
	  $\embed_u$ / $\embed_\graph$ & Node / graph embedding \\ 
      $\targetmodel$ / $\attackmodel$ & Target / attack model \\
      $\propinfermodel$ & Attack model of property inference \\
      $\subinfermodel$ & Attack model of subgraph inference \\
      $\graphreconmodel$ & Attack model of graph reconstruction \\
      \aggr & Aggregation operation \\
      \upd & Updating operation \\
      \pool & Graph pooling operation \\
      \messagee & Message received from neighbors \\
      \subfeat & Feature vector of subgraph inference \\
\bottomrule
\end{tabular}
}
\label{table:notations}
\end{table}

\section{Experimental Details}
\label{app:experimental_details}

\subsection{Graph Sampling Methods}
\label{appsub:detail_graph_sampling}

\begin{itemize}[leftmargin=*]
    \setlength\itemsep{-0.25em}
    \item \mypara{Random Walk Sampling}
    The main idea of \rws is to randomly pick a starting node, and then simulate a random walk on the graph until we obtain the desired number of nodes.
    \item \mypara{Snowball Sampling}
    The main idea of \sbs is to randomly select a set of seed nodes, and then iteratively select a set of neighboring nodes of the selected nodes until we obtain the desired number of nodes.
    \item \mypara{Forest Fire Sampling}
    The main idea of \ffs is to randomly select a seed node, and begin ``burning'' outgoing edges and the corresponding nodes. 
    Here, a node ``burns'' its outgoing edges and the corresponding nodes means these edges and nodes are sampled.
    If an edge gets burned, the node at the other endpoint gets a chance to burn its own edges, and so on recursively until we obtain the desired number of nodes.
\end{itemize}

\begin{figure*}[!t]
\begin{center}
\includegraphics[width=0.8\textwidth]{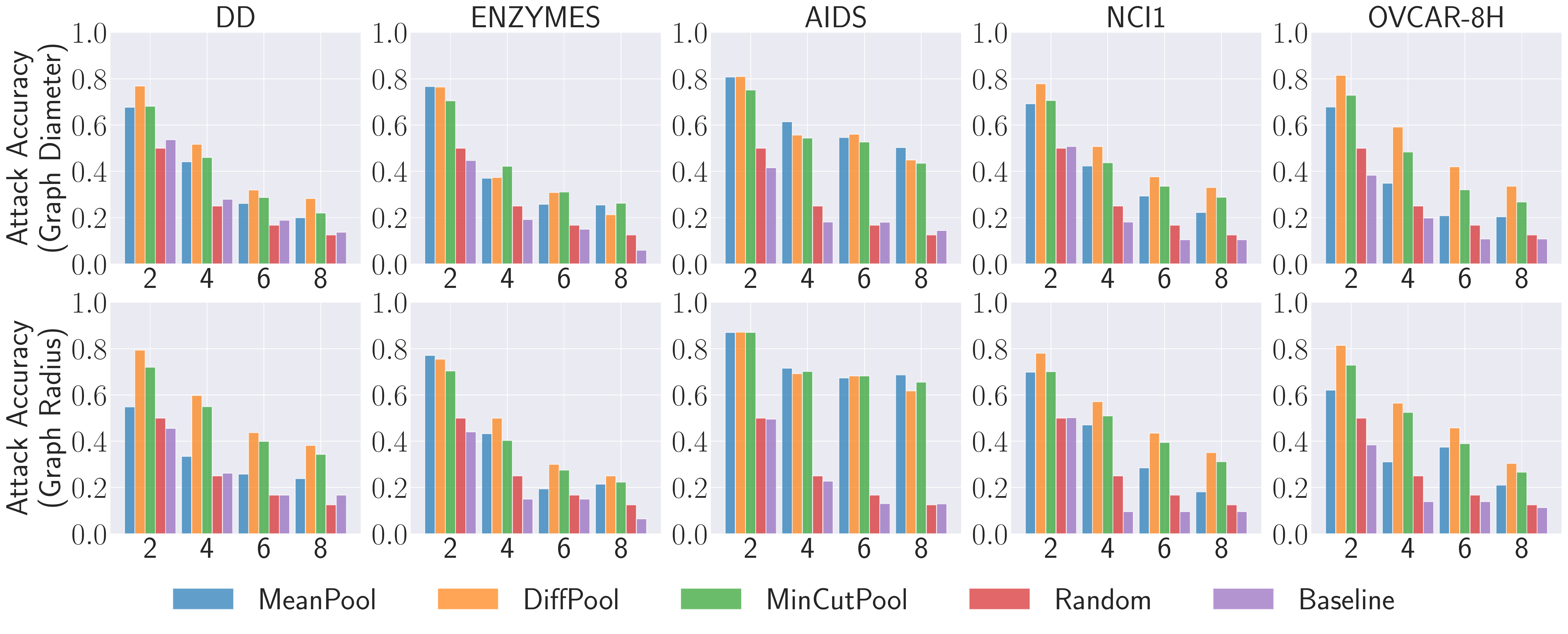}
\end{center}
\vspace{-0.4cm}
\caption{
[Higher means better attack performance.] 
Attack accuracy of additional properties for property inference.
Different columns represent different datasets, and different rows represent different graph properties to be inferred.
In each figure, different legends stand for different graph embedding models, different groups stand for different bucketization schemes.
The $\mathsf{Random}$ and $\mathsf{Baseline}$ method represent the random guessing and summarizing auxiliary dataset baseline, respectively.
}
\label{figure:property_inference_attack2}
\end{figure*}

\begin{figure*}[!th]
\centering
\begin{subfigure}{1.8\columnwidth}
\includegraphics[width=1\columnwidth]{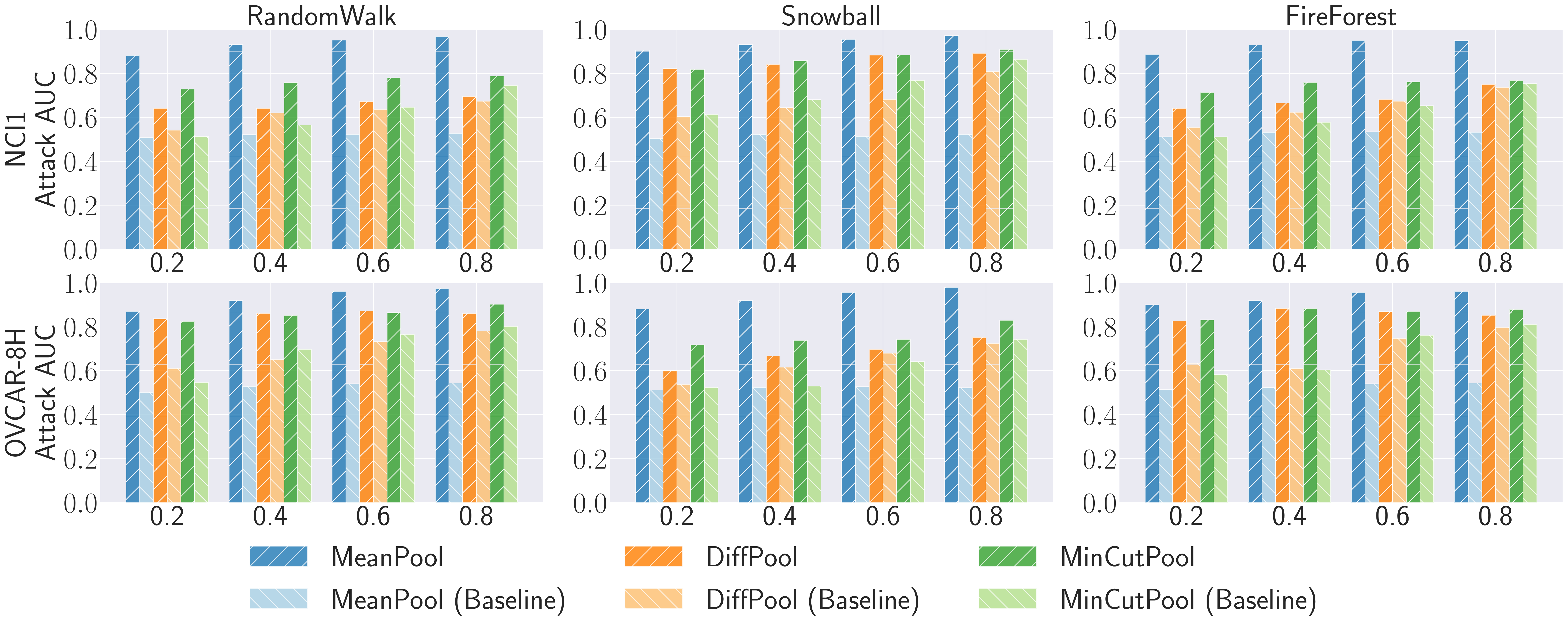}
\end{subfigure} \\
\caption{
[Higher means better attack performance.] 
Attack AUC for subgraph inference attack.
Different rows represent different datasets, and different columns represent different graph sampling methods.
In each figure, different legends stand for different graph embedding models, different groups stand for different sampling ratios.
}
\label{figure:subgraph_inference_attack_compare_2}
\end{figure*} 

\subsection{Macro-level Graph Statistics}
\label{appsub:detail_graph_statistics}

\begin{itemize}[leftmargin=*]
    \setlength\itemsep{-0.25em}
    \item \mypara{Degree Distribution}
    The degree distribution $P(k)$ of a graph is defined to be the fraction of nodes in the graph with degree $k$. 
    It is the most widely used graph statistic to quantify a graph.
    \item \mypara{Local Clustering Coefficient (LCC)}
    The LCC of a node quantifies how close its neighbors are to being a cluster. 
    It is primarily introduced to determine whether a graph is a small-world network.
    \item \mypara{Betweenness Centrality (BC)}
    The betweenness centrality is a measure of centrality in a graph based on the shortest paths. 
    For every pair of nodes in a graph, there exists at least one shortest path between the nodes such that either the number of edges that the path passes through is minimized. 
    The betweenness centrality for each node is the number of these shortest paths that pass through the node.
    \item \mypara{Closeness Centrality (CC)}
    The CC of a node is a measure of centrality in a graph, which is calculated as the reciprocal of the sum of the length of the shortest paths between the node and all other nodes in the graph.
    Intuitively, the more central a node is, the closer it is to all other nodes.
\end{itemize}

\section{Additional Experimental Results}
\label{app:additional_results}

\subsection{Property Inference Attack}
\label{appsub:additional_property_infer}

\mypara{Additional Properties}
\autoref{figure:property_inference_attack2} illustrates the attack performance on the graph diameter and the graph radius properties.
The experimental results show that our attack is still effective on these two properties in most of the settings.
The conclusions are consistent with that of \autoref{subsec:evaluation_properinfer}.

\subsection{Subgraph Inference Attack}
\label{appsub:additional_subgraph_infer}

\mypara{Addtional Datasets}
\autoref{figure:subgraph_inference_attack_compare_2} illustrates the comparison with baseline subgraph inference attacks on the NCI1 and OVCAR-8H datasets.
The conclusions are consistent with that of \autoref{subsec:evaluation_subinfer}.

\begin{figure*}[!ht]
\begin{center}
\includegraphics[width=0.9\textwidth]{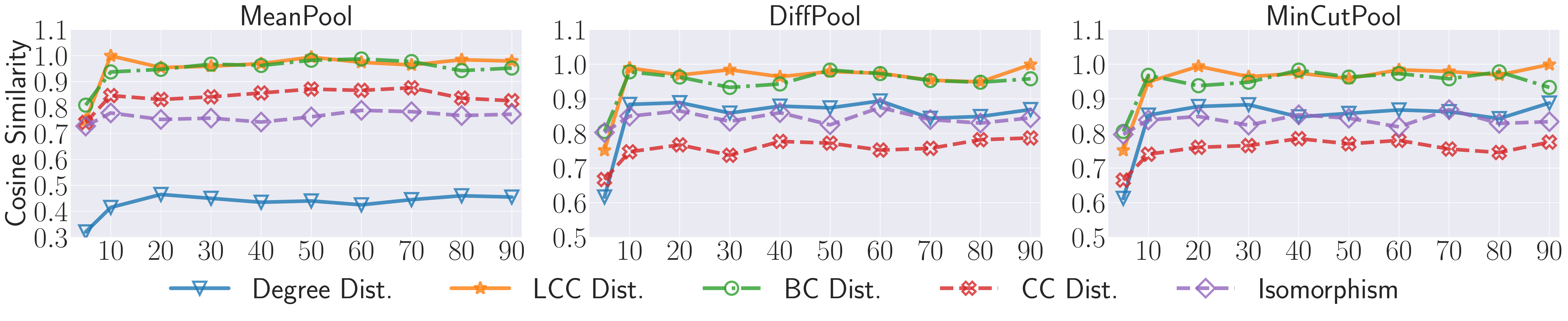}
\end{center}
\vspace{-0.4cm}
\caption{
Impact of the quality of graph auto-encoder on the AIDS dataset.}
\label{figure:graph_reconstruction_ablation}
\end{figure*}

\subsection{Graph Reconstruction Attack}
\label{appsub:additional_graph_recon}

\mypara{Additional Metrics}
\autoref{table:graph_recon_ws} and \autoref{table:graph_recon_js} illustrate the attack performance in terms of macro-level graph statistics measured by Wasserstein distance and JS divergence, respectively.
The experimental results show that our graph reconstruction achieves small Wasserstein distance and JS divergence for most of the settings, indicating our graph reconstruction attack is effective.

\mypara{Impact of Graph Auto-encoder}
To investigate the impact of the quality of the graph auto-encoder on the attack performance, we conduct additional experiments on the graph auto-encoders trained with different epochs.
\autoref{figure:graph_reconstruction_ablation} shows the experimental results.
We observe that with the number of epochs increases, our attack performance increases, indicating the quality of the graph auto-encoder has positive impact on our attack.
When the number of epochs exceeds 10, the attack performance remains unchanged for most of the settings.
Thus, we train the graph auto-encoder for 10 epochs in our experiments.

\begin{figure}[!tpb]
    \centering
    \begin{subfigure}{0.45\columnwidth}
    \includegraphics[width=0.95\textwidth]{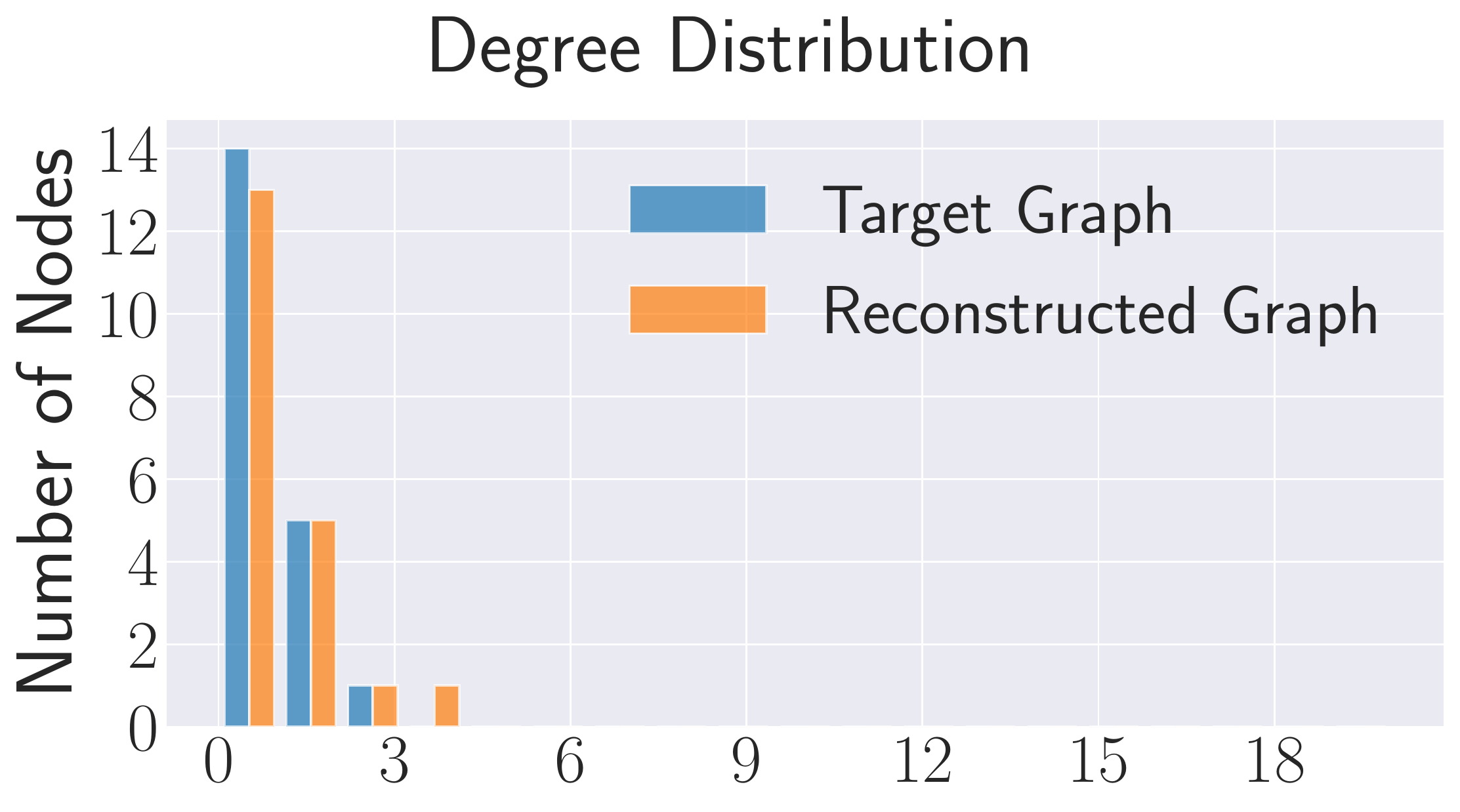}
    \label{subfigure:graph_reconstruction_degree_dist}
    \end{subfigure}
    \begin{subfigure}{0.45\columnwidth}
    \includegraphics[width=0.95\textwidth]{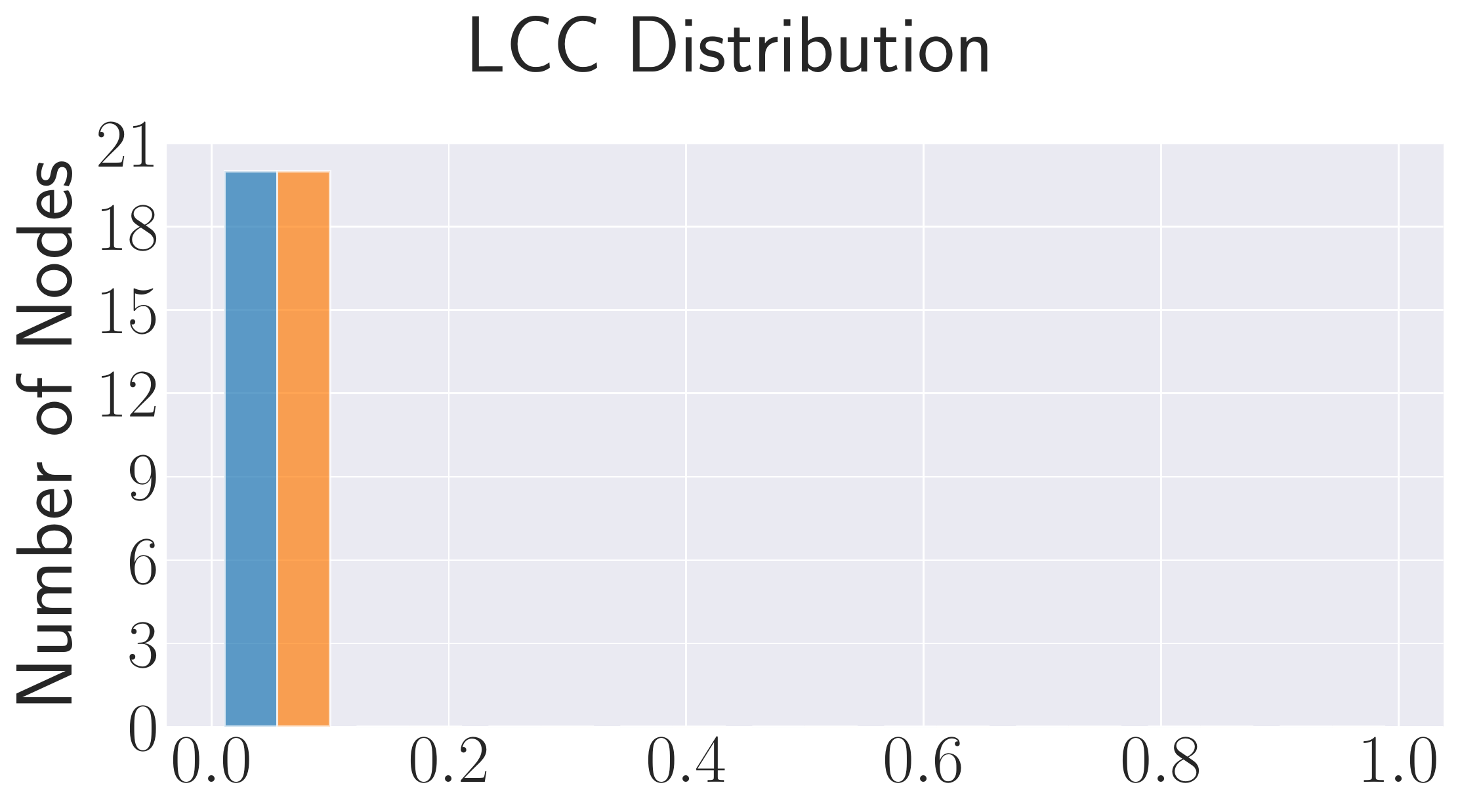}
    \label{subfigure:graph_reconstruction_lcc_dist}
    \end{subfigure} \\
    \begin{subfigure}{0.45\columnwidth}
    \includegraphics[width=0.95\textwidth]{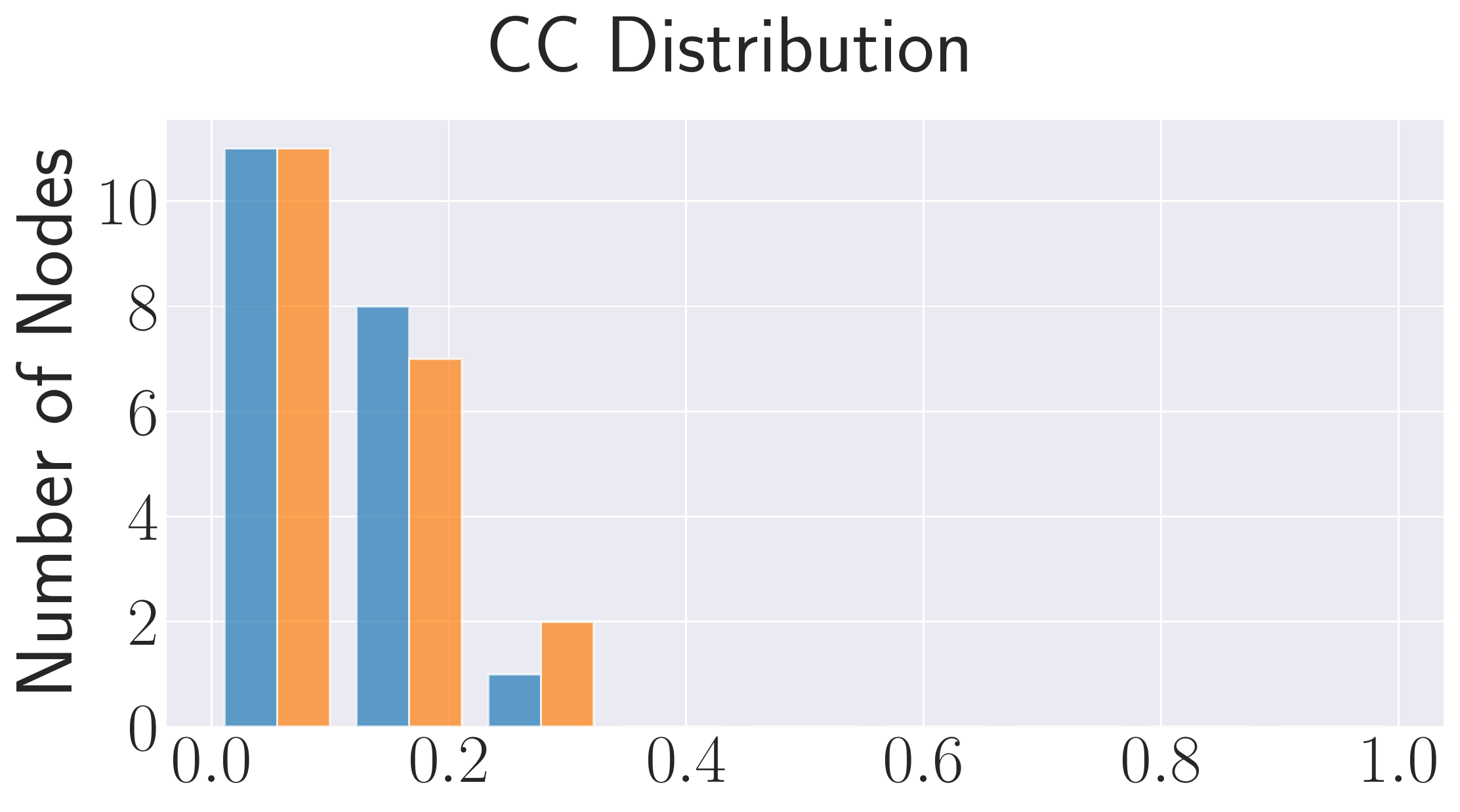}
    \label{subfigure:graph_reconstruction_cc_dist}
    \end{subfigure} 
    \begin{subfigure}{0.45\columnwidth}
    \includegraphics[width=0.95\textwidth]{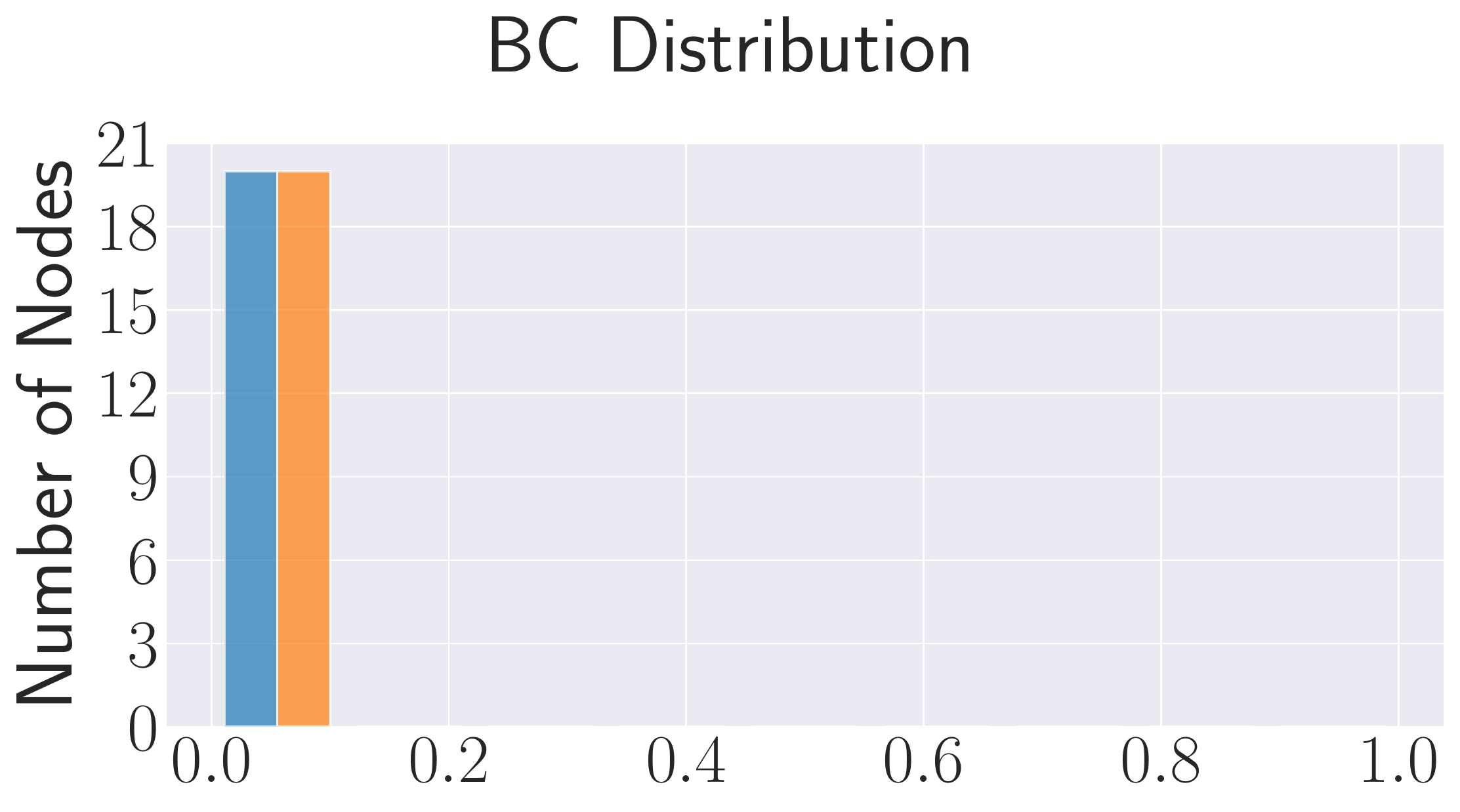} 
    \label{subfigure:graph_reconstruction_bc_dist}
    \end{subfigure}
    \caption{
    Visualization of macro-level graph statistic distribution for graph reconstruction attack on the AIDS dataset.
    }
    \label{figure:graph_recon_visualization}
\end{figure}

\mypara{Visualization}
To better illustrate the effectiveness of our graph reconstruction attack on preserving the macro-level graph statistics, we provide a distribution visualization of the AIDS dataset in \autoref{figure:graph_recon_visualization}.
We experiment on the \mincutpool model.
The visualization results show that our graph reconstruction attack can effectively preserve the macro-level graph statistics.

\begin{table}[!ht]
\centering
\caption{
[Lower means better attack performance.]
Attack performance of graph reconstruction measured by macro-level graph statistics, the similarity of which is measured by Wasserstein distance.
}
\label{table:graph_recon_ws}
\footnotesize
\setlength{\tabcolsep}{0.2em}
\renewcommand{\arraystretch}{1.4}
\scriptsize
\begin{tabular}{c c | c  c  c  c}
\toprule
Dataset & Target Model & \multicolumn{1}{c}{Degree Dist.} & \multicolumn{1}{c}{LCC Dist.} & \multicolumn{1}{c}{BC Dist.} & \multicolumn{1}{c}{CC Dist.}\\
\toprule
\multirow{3}{*}{\rotatebox[origin=c]{90}{AIDS}}
& \diffpool & 0.040 $\pm$ 0.001 & 0.055 $\pm$ 0.002 & 0.011 $\pm$ 0.000 & 0.038 $\pm$ 0.001 \\
& \meanpool & 0.073 $\pm$ 0.000 & 0.020 $\pm$ 0.001 & 0.027 $\pm$ 0.001 & 0.067 $\pm$ 0.001 \\
& \mincutpool & 0.046 $\pm$ 0.000 & 0.067 $\pm$ 0.002 & 0.012 $\pm$ 0.000 & 0.047 $\pm$ 0.001 \\
\midrule
\multirow{3}{*}{\rotatebox[origin=c]{90}{ENZYMES}}
& \diffpool & 0.125 $\pm$ 0.004 & 0.201 $\pm$ 0.009 & 0.039 $\pm$ 0.001 & 0.258 $\pm$ 0.005 \\
& \meanpool & 0.060 $\pm$ 0.006 & 0.188 $\pm$ 0.018 & 0.039 $\pm$ 0.001 & 0.086 $\pm$ 0.009 \\
& \mincutpool & 0.085 $\pm$ 0.006 & 0.199 $\pm$ 0.005 & 0.040 $\pm$ 0.003 & 0.171 $\pm$ 0.013 \\
\midrule
\multirow{3}{*}{\rotatebox[origin=c]{90}{NCI1}}
& \diffpool & 0.063 $\pm$ 0.001 & 0.091 $\pm$ 0.004 & 0.056 $\pm$ 0.001 & 0.084 $\pm$ 0.003 \\
& \meanpool & 0.045 $\pm$ 0.001 & 0.049 $\pm$ 0.004 & 0.062 $\pm$ 0.001 & 0.067 $\pm$ 0.001 \\
& \mincutpool & 0.087 $\pm$ 0.000 & 0.119 $\pm$ 0.003 & 0.055 $\pm$ 0.001 & 0.138 $\pm$ 0.001 \\
\bottomrule
\end{tabular}
\end{table}

\begin{table}[!ht]
\centering
\caption{
[Lower means better attack performance.]
Attack performance of graph reconstruction measured by macro-level graph statistics, the similarity of which is measured by JS divergence.
}
\label{table:graph_recon_js}
\footnotesize
\setlength{\tabcolsep}{0.2em}
\renewcommand{\arraystretch}{1.4}
\scriptsize
\begin{tabular}{c c | c  c  c  c}
\toprule
Dataset & Target Model & \multicolumn{1}{c}{Degree Dist.} & \multicolumn{1}{c}{LCC Dist.} & \multicolumn{1}{c}{BC Dist.} & \multicolumn{1}{c}{CC Dist.}\\
\toprule
\multirow{3}{*}{\rotatebox[origin=c]{90}{AIDS}}
& \diffpool & 0.120 $\pm$ 0.003 & 0.052 $\pm$ 0.002 & 0.029 $\pm$ 0.001 & 0.080 $\pm$ 0.005 \\
& \meanpool & 0.253 $\pm$ 0.001 & 0.019 $\pm$ 0.000 & 0.056 $\pm$ 0.002 & 0.132 $\pm$ 0.004 \\
& \mincutpool & 0.136 $\pm$ 0.000 & 0.068 $\pm$ 0.003 & 0.029 $\pm$ 0.001 & 0.106 $\pm$ 0.001 \\
\midrule
\multirow{3}{*}{\rotatebox[origin=c]{90}{ENZYMES}}
& \diffpool & 0.341 $\pm$ 0.007 & 0.279 $\pm$ 0.012 & 0.071 $\pm$ 0.006 & 0.540 $\pm$ 0.014 \\
& \meanpool & 0.201 $\pm$ 0.015 & 0.213 $\pm$ 0.009 & 0.073 $\pm$ 0.003 & 0.165 $\pm$ 0.019 \\
& \mincutpool & 0.280 $\pm$ 0.004 & 0.248 $\pm$ 0.003 & 0.073 $\pm$ 0.006 & 0.354 $\pm$ 0.028 \\
\midrule
\multirow{3}{*}{\rotatebox[origin=c]{90}{NCI1}}
& \diffpool & 0.210 $\pm$ 0.001 & 0.103 $\pm$ 0.002 & 0.093 $\pm$ 0.003 & 0.206 $\pm$ 0.006 \\
& \meanpool & 0.159 $\pm$ 0.004 & 0.048 $\pm$ 0.003 & 0.105 $\pm$ 0.001 & 0.149 $\pm$ 0.003 \\
& \mincutpool & 0.275 $\pm$ 0.000 & 0.160 $\pm$ 0.003 & 0.085 $\pm$ 0.001 & 0.345 $\pm$ 0.005 \\
\bottomrule
\end{tabular}
\end{table}

\subsection{Defense}
\label{appsub:additional_defense}

\begin{figure*}[!bp]
\begin{center}
\includegraphics[width=0.9\textwidth]{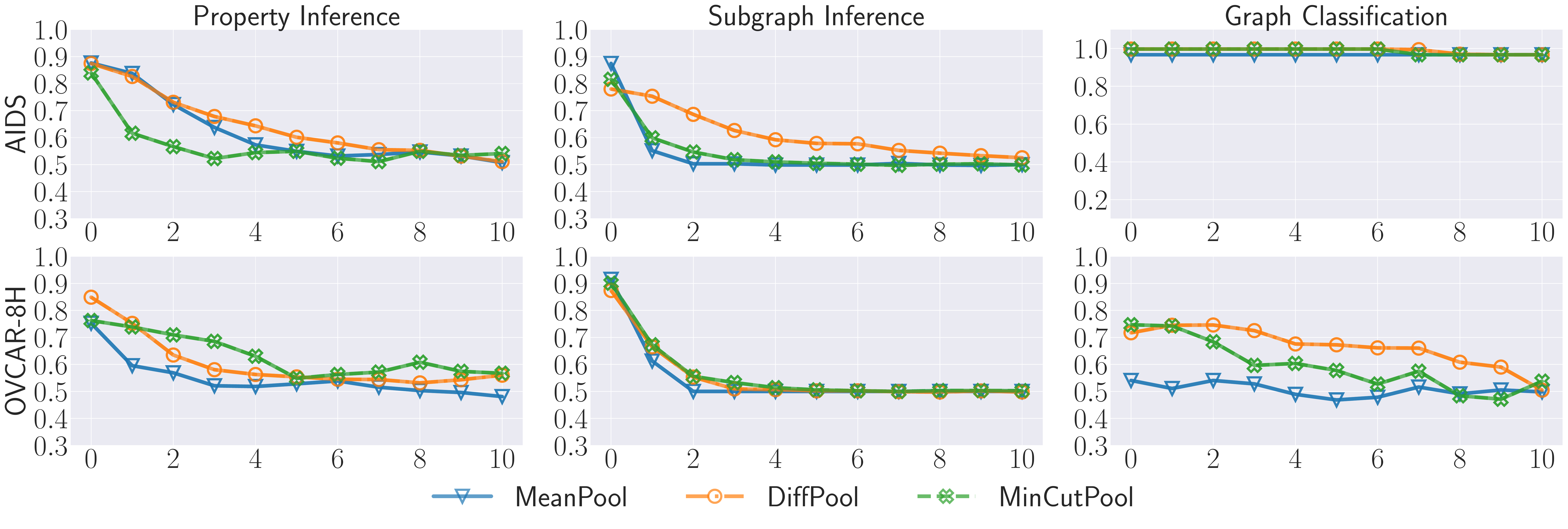}
\end{center}
\vspace{-0.4cm}
\caption{
Graph embedding perturbation defense on the AIDS and OVCAR-8H datasets.
The first and second column represents the attack performance of property inference and subgraph inference respectively, the last column represents the accuracy of normal graph classification task.
In each figure, the x-axis stands for the scaling parameter $\beta$ for Laplace noise, where larger $\beta$ means higher noise level.
The y-axis stands for the attack performance/normal graph classification accuracy.
}
\label{figure:defense_2}
\end{figure*}

\mypara{Additional Datasets}
\autoref{figure:defense_2} illustrates the defense performance on the ADIS and OVCAR-8H datasets for property inference and subgraph inference attack.
The conclusions are consistent with that of \autoref{sec:defense} for these datasets.

\begin{figure*}[!pb]
\begin{center}
\includegraphics[width=0.9\textwidth]{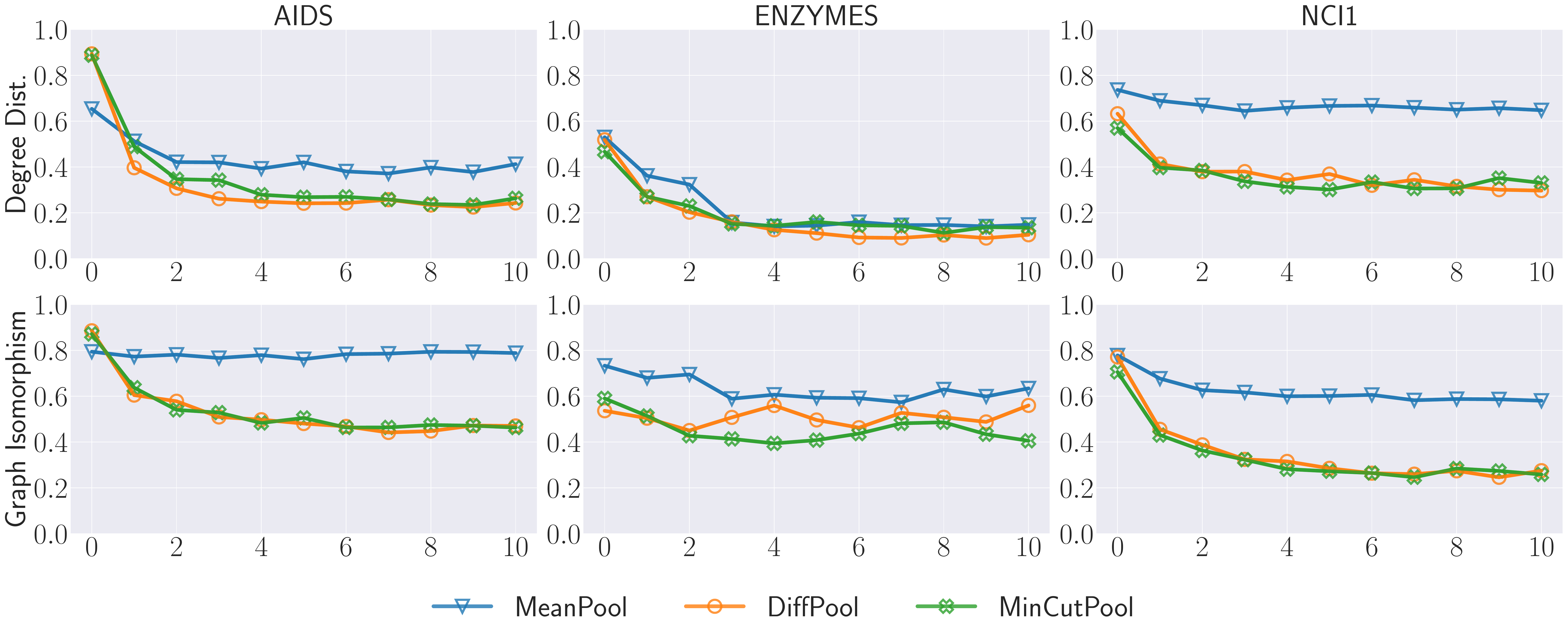}
\end{center}
\vspace{-0.4cm}
\caption{
Graph embedding perturbation defense against the graph reconstruction attack.
In each figure, the x-axis stands for the scaling parameter $\beta$ for Laplace noise, where larger $\beta$ means higher noise level.
The y-axis stands for the cosine similarity of degree distribution and graph isomorphism, respectively.
}
\label{figure:defense_graph_recon}
\end{figure*}

\mypara{Defense against Graph Reconstruction}
\autoref{figure:defense_graph_recon} illustrates the defense performance for graph reconstruction attack.
The experimental results show that our defense mechanism is still effective for graph reconstruction attack.

\section{Impact of Node Features}
\label{app:impact_node_feature}
\begin{figure*}[!htpb]
\begin{center}
\includegraphics[width=\textwidth]{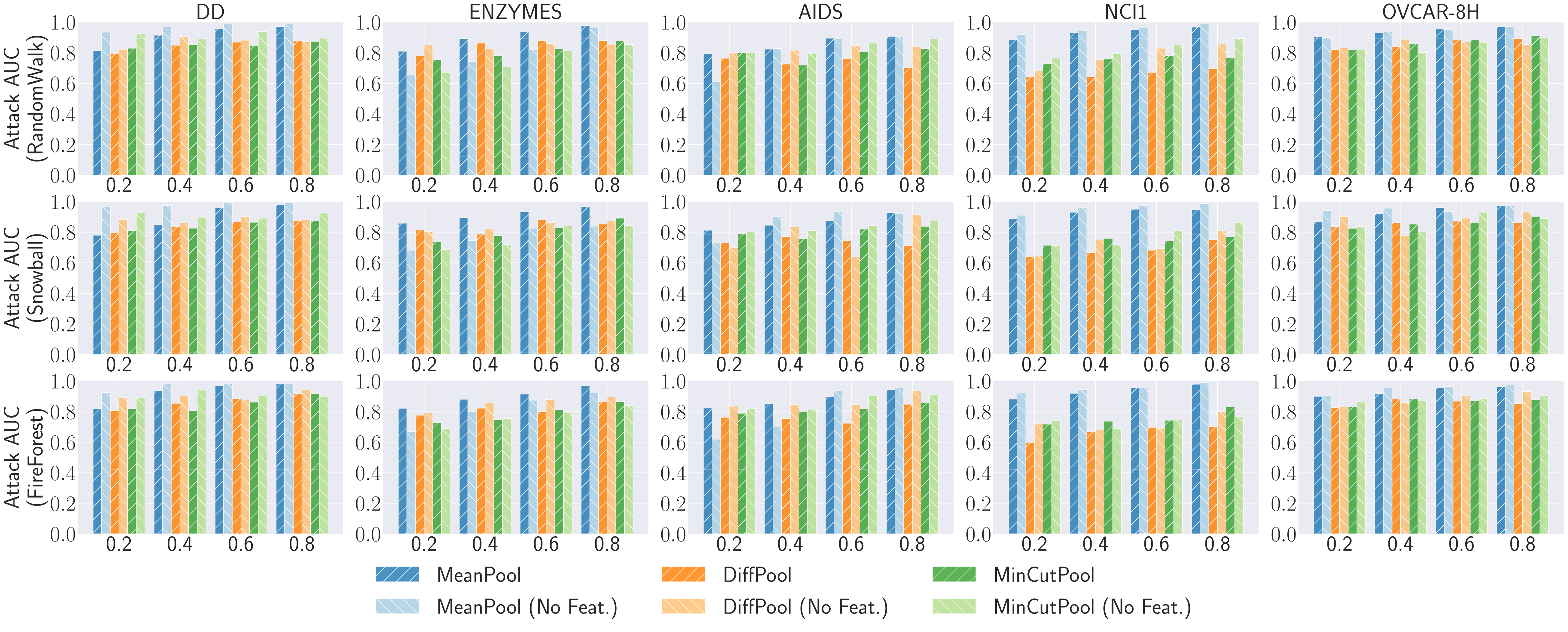}
\end{center}
\vspace{-0.4cm}
\caption{
Comparison of attack AUC between graphs with and without node features for subgraph inference attack.
}
\label{figure:subgraph_inference_auc_no_feat}
\end{figure*}

To evaluate the impact of node features, we conduct additional experiments on graphs without node features.
Concretely, for each dataset in~\autoref{table:dataset_statistic}, we replace all its original node features with one-hot encodings of node degrees.
This follows the setting of~\cite{XHLJ19} which aims to investigate the expressiveness of graph structure.
\autoref{figure:subgraph_inference_auc_no_feat} shows the experimental results for the subgraph inference attack.
The experimental results show that the attack performance of graphs with and without node features is similar for most of the settings, indicating the robustness of our subgraph inference attack.

\end{document}